\documentstyle[11pt]{article}
\def\up#1{\raise 1ex\hbox{#1}}
\begin{document}
\centerline{\bf MARTINGALE PROBLEM APPROACH TO THE REPRESENTATIONS}
\centerline{\bf OF THE NAVIER-STOKES EQUATIONS ON SMOOTH MANIFOLDS}
\centerline{\bf  WITH SMOOTH BOUNDARY}

\bigskip
\bigskip
\centerline{\bf Diego L. Rapoport}
\bigskip
\centerline{Applied Mechanics,FIUBA, Univ. of Buenos Aires, P.Col\'on 850}
\centerline {Buenos Aires 1069, Argentina; drapo@unq.edu.ar.}

\bigskip

{\bf ASTRACT:} We present the random representations for the Navier-Stokes vorticity
equations
for an incompressible fluid in a smooth manifold with smooth boundary and reflecting boundary conditions
for the vorticity. We specialize our constructions to $R^{n-1}\times R_+$. We extend these constructions to give
the random representations for the kinematic dynamo problem of magnetohydrodynamics. We carry out these integrations through the
application of the methods of Stochastic Differential Geometry, i.e. the gauge theory
of diffusion processes on smooth manifolds. 


{\bf AMS SUBJECT CLASSIFICATION}: 35Q30, 60H10, 60H30, 60J60, 76D06, 76M35.

\bigskip

\section{Introduction.}

\hspace{2em} 
Statistical approaches to the equations of fluid dynamics are well known [10,11].
The present author, following a different stochastic approach to the previously
quoted classical works, has given random implicit representations for the Navier-Stokes equations for an incompressible
fluid on a smooth compact manifold without boundary. The approach chosen for the obtention of these representations stemmed
from Stochastic Differential Geometry, i.e. the invariant theory of diffusion processes on smooth manifolds, and in particular, in Euclidean space 
 [1,2,8,9,14,23-27];
this approach has lead the author to give
as well random representations for the kinematic dynamo problem, i.e. the equations of transport
of a passive magnetic field transported by the fluid, in the same geometric situation [5].  While stemming from
the fussion of gauge theory and stochastic processes on manifolds, the present theory
is founded on the relation between a class of linear connections with torsion [13,23,27], and
the random diffeomorphisms determined by them, extending thus the classical diffeomorphims 
associated to perfect fluids, presented in the pioneering work of V.I. Arnold [15,16]
and further ellaborated by Ebin and Marsden [17]. In a recent interesting program,
Gliklikh [18-20] has extended the original
proposal of Arnold-Ebin-Marsden, to give a formulation of the Navier-Stokes equations on the $n$-torus 
as a stochastic
perturbation of the group of volume preserving diffeomorphims associated to the flow of perfect flows. 
In the present gauge-theoretical approach, the representations obtained by the present author were
obtained in two instances [5,6,12,13]: Firstly, using the so-called derived (or jacobian) random flow, constructed as the derivative flow
of the lagrangian random representations for the fluid particles, in case the manifold is isometrically
immersed in Euclidean space (the case of gradient diffusion processes). Secondly, for the more general case of arbitrary
compact manifolds, the representations were derived in terms of the Hessian and Ricci random flows related to the former processes.
In this note we wish to extend these constructions of the random representations for the vorticity Navier-Stokes 
and kinematic dynamo equations,
to the case of a smooth manifold with smooth boundary, both for which no general representations are known; in this case, the boundary
conditions for the vorticity in practice in Fluid-dynamics, will turn to be the well known reflection at the boundary conditions which we shall reintroduce below.
Yet, our presentation will follow instead the original approach to Brownian motion on smooth manifolds
which lead to Stochastic Differential Geometry: The stochastic extension of the Cartan classical soldering method, as originally
presented by Malliavin [8] and Elworthy [2], which we have presented as the mathematical approach
to Statistical Thermodynamics, Quantum Mechanics and Gravitation [4,13,23-27,29], while the classical method appeared to be adequate
for the formulation of the dynamics of classical relativistic spinning particles subjected to gravitational fields [22,28] without 
using no lagrangian nor hamiltonean structures. In this setting, the torsion is no longer provided by the full irreducible expression
of the torsion tensor, instead, the trace-torsion is the basic 'physical' field to be taken in account in a theory of generalized Brownian motions, since
its conjugate vector-field turns to be describe the drift of the process. Yet, while in the setting of classical differential geometry in which the
transport of fields is done along smooth curves so that the Cartan calculus of differential forms is the basic tool, in this continuous 
non-differentiable setting this calculus has
to be supplemented by the rules of stochastic analysis. These together yield a most formidable theoretical and computational instrument, and the Ito formula
for differential forms, its highest and most basic expression of this extended calculus. Thus, it will turn to be of no surprise, that it is precisely this formula which will
give us the key to the derivation of the analytical expression for the evolution Navier-Stokes equations for the vorticity on a manifold with smooth
boundary. This approach will turn out to extend not only the analytical expressions obtained in the case of no boundary case obtained by the author, but
the original approach in Computational  Fluid-dynamics known as the random vortex method, which deals with viscous fluids on the Euclidean plane [40].

\section{Riemann-Cartan-Weyl Connections and Their Laplacians}
\hspace{2em}
We shall assume an $n$-dimensional smooth connected oriented compact manifold, $M$. 
We shall further assume as given a Riemann-Cartan-Weyl connection on $M$, $\nabla$, i.e. a linear connection on $M$ (or still, a covariant derivative on
$TM$) 
which is $g$-compatible,  and such that its
Christoffel coefficients defined by $\nabla_{{\partial \over \partial x^\beta}}{\partial \over \partial x^\gamma}
 = \Gamma^\alpha_{\beta \gamma}{\partial \over \partial x^\alpha}$
are of the form
\begin{eqnarray}
\Gamma ^{\alpha}_{\beta \gamma}(x) = {\alpha\brace\beta\gamma}(x)~+~ {2 \over (n-1)}
\left\{\delta^\alpha_\beta ~Q_\gamma (x)~ - ~ g_{\beta\gamma}(x)~Q^\alpha (x)\right\}, x\in M
\end{eqnarray} 
with torsion tensor 
\begin{eqnarray*}
T^{\alpha}_{\beta \gamma}(x) = 1/2(\Gamma^\alpha _{\beta \gamma}(x) - \Gamma^{\alpha}_{\gamma \beta}(x)),
\end{eqnarray*}
which has an irreducible decomposition which is zero for all its components with exception of the trace-torsion
$1$-form given by 
\begin{eqnarray*}
Q(x)= Q_\beta(x)dx^\beta:= T^{\alpha}_{\alpha \beta}(x)dx^\beta. 
\end{eqnarray*}
The first term in $(1)$ stands for the Christoffel coefficients of the unique torsionless
linear connection determined a Riemannian metric $g$, i.e. the Levi-Civita connection, which we shall denote in the following as $\nabla^ g$.

Associated to this connection we have a Laplacian operator defined by 
\begin{eqnarray}
H_0(g,Q): = 1/2 \nabla^2
\end{eqnarray}
(the subscript $0$ denotes its actions on functions, which we here consider to be smooth, as well as all other fields that appear subsequently)
which only depends on $g$, its derivatives, and $Q$, since one can compute explicitly to obtain
\begin{eqnarray}
H_0(g,Q) = 1/2 \triangle_g + L_{Q^g}.
\end{eqnarray}
In this expression
\begin{eqnarray*}
\triangle_ g = {\rm div~} {\rm grad} = (\nabla^g)^2
\end{eqnarray*}
is the Laplace-Beltrami operator
associated to $\nabla^g$.
Furthermore, $Q^g$ is the vectorfield conjugate to $Q$ by $g$, i.e.$Q^g (f) = g(Q,df)$, for any smooth function $f$,
and $L_X$ denotes the Lie-derivative
with respect to the vectorfield $X$ on $M$. Recall that on functions $f$ we have the identity $L_{Q^g}f = Q^g(f)$, so that by recalling that the metric is non-degenerate, we conclude
that expression $(3)$ corresponds to that of the most general laplacian with zero-potential term, acting on functions. We note that the second term in expression
$(1)$ describes the drift, and appears that in an invariant setting, it is related to the trace-torsion $1$-form.

These operators can be rewritten in a form suitable to carry out stochastic analysis on $M$ and on $P_{O(n)}$, the bundle of orthogonal frames 
of the tangent space $TM$ over $M$ [1,30], i.e.
\begin{eqnarray*}
P_{O(n)} = \{r = (x,e(x)), x \in M, e(x) = e^\alpha _a (x){\partial \over \partial x^\alpha}|_x  {\rm ~a~ basis~of~}T_xM\}  
\end{eqnarray*}
so that
\begin{eqnarray*}
g_{\alpha \beta}e^\alpha _a e^\beta _a = \delta_{ab}.
\end{eqnarray*}
Consider the canonical horizontal vectorfields on $P_{O(n)}$, $L^\nabla$ and $L^{\nabla^g}$
defined by the connections $\nabla$ and $\nabla^g$ respectively, with components $L^\nabla _a, L^{\nabla^g}_a, a=1,\ldots,n$ respectively, given
by \footnote{Einstein repeated sign sum convention is assumed in the following.}
\begin{eqnarray}
L^\nabla_a F(r) = e^\alpha_a (x){\partial F \over \partial x^\alpha}(r) - \Gamma^\alpha_{\beta \gamma}(x)
e^\gamma_a (x)e^\beta_c (x){\partial F\over \partial e^\alpha _c}(r), a=1,\ldots,n
\end{eqnarray}
and
\begin{eqnarray}
L^{\nabla^g}_ a F(r) = e^\alpha_a (x){\partial F \over \partial x^\alpha}(r) -{\alpha\brace\beta\gamma}(x)e^\gamma_a (x)e^\beta_c (x){\partial F \over \partial
e^\alpha _c}(r), a=1,\ldots,n.
\end{eqnarray}
Then, the horizontal lift of the operator $H_0(g,Q)$ acting on functions
defined on $P_{O(n)}$, which we denote as $\tilde H_0(g,Q)$ is (see Theorem $1.2$, page $238$ in [1])
\begin{eqnarray}
\tilde H_0(g,Q) = {1\over 2} L^\nabla_a(L^\nabla_a),
\end{eqnarray}
so that for any "basic" function, $\tilde f(r) =f\circ \pi (r)\equiv f(x),r=(x,e)$, 
where $\pi:P_{O(n)}\rightarrow M$, 
$\pi(r):=x$ for any $r=(x,e)\in P_{O(n)}$ is the bundle projection, we have the identity
\begin{eqnarray}
[\tilde H_0(g,Q)\tilde f](r) = [H_0(g,Q)f](x)
\end{eqnarray}
and the horizontal lift of the Laplace-Beltrami operator, $\tilde \triangle _g$ consequently
verifies 
\begin{eqnarray}
\tilde \triangle_g = 2\tilde H_0(g,0) =  L^{\nabla^g}_a (L^{\nabla^g}_a).
\end{eqnarray}
Note that we also have 
\begin{eqnarray}
\tilde H_0(g,Q) = {1\over 2}(L^{\nabla^g}_a(L^{\nabla^g}_a))+ L_{\tilde Q^g},
\end{eqnarray}
where $\tilde Q^g$ is the horizontal lift of $Q^g$, i.e. $\tilde Q^g (\tilde f):= \tilde (Q^g(f))\equiv \pi^*(Q^g(f))$.

We can now extend these operators to act on smooth differential forms defined on $M$ and its horizontal lifts to act on smooth
differential forms on $P_{O(n)}$. Indeed, consider
\begin{eqnarray}
H_p(g,Q) = {1\over 2}\triangle_p + L_{Q^g} |_{\Lambda^p (T^*M)},
\end{eqnarray}
where
\begin{eqnarray*}
\triangle_p = (d-\delta)^2|_{(\Lambda^p (T^* M))}
\end{eqnarray*}
is the Hodge laplacian acting on differential $p$-forms defined on $M$, so that $d$ and $\delta$ are the exterior differential and
codifferential operators, respectively.  We recall that $\delta$ is the adjoint operator to $d$ with respect to the pairing  
\begin{eqnarray*}
<<\phi,\nu>> = \int \phi^{\alpha_1 \ldots \alpha_p}(x)\nu_{\alpha_1\ldots \alpha_p}(x)vol_g(x),
\end{eqnarray*}
where $vol_g(x)= {\rm det}~(g)dx^1\wedge \ldots \wedge dx^n$ is the Riemannian volume density,
$\phi = {1\over p!}\phi_{\alpha_1 \ldots \alpha_p}(x)dx^{\alpha_1}\wedge \ldots \wedge dx^{\alpha_p}$ and $\nu={1\over p!}\nu_{\alpha_1\wedge \ldots
\wedge \alpha_p}(x)dx^{\alpha_1}\wedge \ldots \wedge dx^{\alpha_p}$ are $p$-forms and $\phi^{\alpha_1 \ldots \alpha_n} = g^{\beta_1 \alpha_1}\ldots
g^{\beta_p\alpha_p}\phi_{\beta_1 \ldots \beta_p}$. Since $d^2=0$, then also $\delta^2 = 0$ and consequently
\begin{eqnarray}
\triangle_p = -(d\delta + \delta d)|_{\Lambda^p (T^*M)}
\end{eqnarray}
and then
\begin{eqnarray*}
H_p(g,Q) = -{1 \over 2}(\delta d + d\delta) + L_{Q^g} |_{\Lambda^p (T^*M)}.
\end{eqnarray*}
As well known, the Lie-derivative of a $p$-differential form with respect to a vectorfield
is well
defined and independant of the metric $g$:
\begin{eqnarray*}
L_X \omega = (i_X d + di_X) \omega
\end{eqnarray*}
where $i_X$ is the interior derivative with respect to the vector field $X$ and $\omega$ is an arbitrary $p$-form on $M$
, and in case of $p= 0$, since $i_X f = 0$ for an arbitrary scalar field $f$ ($i_X$ reduces degree) we get the original laplacian in $(3)$. 
While the second term is independant of the metric, the Hodge laplacian depends on the metric and the curvature obtained by taking
appropiate expressions on its derivatives. 
For the forthcoming discussion it is essential we display the dependance of the Hodge laplacian on the curvature, to write the whole
laplacian of our interest given by $(10)$ in the form
\begin{eqnarray*}
(H_p(g,Q)\phi)_{\alpha_1 \ldots \alpha_p} = {1 \over 2}(\triangle_g \phi)_{\alpha_1\ldots \alpha_p}
    -{1 \over 2}(-1)^\nu \sum_{\nu = 1}^p R^{\beta.\gamma.}_{.\gamma .\alpha_\nu}\phi_{\beta\alpha_1\ldots\hat\alpha_\nu \ldots \alpha_p}
\end{eqnarray*}
\begin{eqnarray}
- \sum_{1\leq \mu < \nu \leq p}(-1)^{\nu + \mu}R^{\beta.\gamma .}_{.\alpha_\nu . \alpha_\mu}
\phi_{\gamma \beta \alpha_1 \ldots \hat\alpha_\mu \ldots \hat\alpha_\nu \ldots \alpha_p}
 + (L_{Q^g}\phi)_{\alpha_1 \ldots \alpha_p}.
\end{eqnarray}
Here, as usual, $\hat \alpha$ denotes ommision of the index $\alpha$.
The second and third terms in the r.h.s. of $(12)$, constitute the well known Weitzenbock term, which of course, vanishes completely
in the case $p=0$. Note that for $2$-forms, 
\begin{eqnarray*}
\Omega = {1 \over 2}\Omega_{\alpha_1 \alpha_2}dx^{\alpha_1} \wedge dx^{\alpha_2}, 
\end{eqnarray*}
the Weitenzbock term is
\begin{eqnarray}
-{1\over 2}(-1)^\nu \sum_{\nu = 1}^2 R^\beta_{\alpha_\nu} \Omega_{\beta \alpha_\nu}
-\sum_{\mu < \nu}(-1)^{\nu + \mu}R^{\beta.\gamma.}_{.\alpha_\nu . \alpha_\mu}\Omega_{\gamma \beta} 
\end{eqnarray}
where in the first term we have a coupling of the differential form to the Ricci curvature tensor, ${\rm Ric} = (R^\beta_{\alpha}) =
(R^{\beta.\gamma.}_{.\gamma.\alpha})$, and 
$R = (R^{\alpha .\beta .}_{.\gamma.\delta})$ is the $(2,2)$
tensor associated to the Riemannian curvature tensor [31].

Now, given a $p$-form defined on $M$, we can define a $p$-form on $P_{O(n)}$, its horizontal lift, and thus further establish an isomorphism between
$\Lambda^p (T^*M)$ and
${\rm hor}~(\Lambda^p (T^*P_{O(n)}))$, the horizontal $p$-forms defined on $P_{O(n)}$, i.e. the subspace of $\Lambda^p(T^*P_{O(n)})$ whose projection
by the bundle mapping $\pi$ has zero kernel, i.e. $\pi^*({\rm hor}~(\Lambda^p T^*P_{O(n)})) = \Lambda^p(T^*M)$. Indeed, consider
the co-frame bundle $P^*_{O(n)} =\{(x,\theta (x)= (\theta^ a(x))=(\theta^a_\alpha (x)dx^ \alpha \in T^*_x
M), \theta^a(x)(e_b)(x) = \delta^a_b, {\rm for~any~} x\in M\}$, so that $\theta^a_\alpha \theta^b_\beta g^{\alpha \beta} = \delta _{ab}$. Then, given a
$p$-form $\Omega $ on $M$,
\begin{eqnarray}
\Omega = \Omega (x)= {1 \over p!}\Omega_{\alpha_1 \ldots \alpha_p}(x)dx^{\alpha_1}\wedge \ldots \wedge dx^{\alpha_p},
\end{eqnarray}
we can define functions
on $P_{O(n)}$ given by 
\begin{eqnarray}
\tilde \Omega _{a_1\ldots a_p}(r):= \Omega_{\alpha_1 \ldots\alpha_p}(x)e^{\alpha_1}_{a_1}(x)\ldots e^{\alpha_p}_{a_p}(x), r=(x,e)
\end{eqnarray}
and thus obtain
a (horizontal) $p$-form on $P_{O(n)}$, i.e. an element of ${\rm hor}(\Lambda^p(T^*P_{O(n)}))$ defined by
\begin{eqnarray}
\tilde \Omega = \tilde \Omega(r) = {1 \over p!} \Omega_{a_1 \ldots a_p}(r)\theta^{a_1}\wedge\ldots \wedge\theta^{a_p}.
\end{eqnarray}
This definition can be inverted: if we start with $\tilde \Omega \in
\Lambda^p T^*(P_{O(n)})$, we can obtain the $p$-form $\Omega$ defined on $M$ by
\begin{eqnarray}
\Omega = \Omega(x)  = {1 \over p!}\Omega_{\alpha_1 \ldots \alpha_p}(x) dx^{\alpha_1}\wedge \ldots \wedge dx^{\alpha_p}, 
\end{eqnarray}
where
\begin{eqnarray*}
\Omega_{\alpha_1\ldots \alpha_p}(x) & = & \tilde \Omega_{a_1 \ldots a_p}(r)\theta^{a_1}_{\alpha_1}(x)\ldots \theta^{a_p}_{\alpha_p}(x), r=(x,e), \theta = e^{-1}.
\end{eqnarray*}
Naturally, $\tilde \Omega$ defined by $(15)$
is $O(n)$-equivariant, i.e. 
\begin{eqnarray}
\tilde \Omega_{a_1 \ldots a_p}(r) = \tilde \Omega_{b_1\ldots b_p}(T_Ar)B^{b_1}_{a_1}\ldots B^{b_p}_{a_p}, 
\end{eqnarray}
for any $r \in P_{O(n)}$, $A \in P_{O(n)}$,
$B$ the inverse of $A$ and $T_Ar = Ar$.

Suppose given a $p$-form $\tilde \omega$ defined on $P_{O(n)}$; then, we can consider a time-dependant $p$-form $\tilde \Omega$ defined on
$[0,\infty)\times P_{O(n)}$, by the expression
\begin{eqnarray}
\tilde \Omega_{a_1 \ldots a_p}(\tau,r) = E_r[\tilde \omega _{a_1 \ldots a_p}(r(\tau,r,w))],
\end{eqnarray}
where $E$ denotes expectation 
value,   $w:[0,\infty) \rightarrow R^n$ is a continuous mapping such that $w(0) = 0$ is the canonical realization of a Wiener process,   
and $r =r(\tau,r,w)$ satisfies the Stratonovich stochastic differential equation
\begin{eqnarray}
dr(\tau) = L^\nabla_a(r(\tau))\circ dw(\tau), r(0)= r.
\end{eqnarray}
We note that if $r(0) \in P_{O(n)}$, then $r(\tau, r,w)\in P_{O(n)},\forall \tau \geq 0$. 
Thus defined, the scalar components $\tilde \Omega_{a_1 \ldots a_p}$ of $\tilde \Omega(\tau,r) ={1\over p!}\tilde \Omega_{a_1\ldots a_p}\theta^{a_1}\wedge \ldots \wedge \theta^{a_p}
\in \Lambda^p([0,\infty)\times T^*P_{O(n)})$, ($1 \leq a_1< \ldots <a_p \leq p$)are the unique solution of the Cauchy problem (see page $286$ in [1])
\begin{eqnarray}
{\partial V \over \partial \tau} = \tilde H_0(g,Q)V
\end{eqnarray}
with
\begin{eqnarray}
V(0,-) = \tilde \omega_{a_1 \ldots a_p}, \forall 1\le a_1 <\ldots <a_p \leq p.
\end{eqnarray}

We wish to solve, more generally, the Cauchy problem for $p$-forms on $[0,\infty)\times P_{O(n)}$:
\begin{eqnarray}
{\partial \tilde \Omega \over \partial \tau}(\tau,r) = \tilde H_p(g,Q)\tilde \Omega (\tau,r),
\end{eqnarray}
with given
\begin{eqnarray}
\tilde \Omega (0,-) = \tilde \omega \in \Lambda^pT^*P_{O(n)},
\end{eqnarray}
where we notice that in distinction with the problem $(21 \& 22)$, a coupling of the Weitzenbock term to the components of $\tilde \Omega$ is to be accounted.
This problem is simply solved through the use of the Feynman-Kac formula. For this we consider the canonical realization of the generalized horizontal Brownian motion, given by
the set 
\begin{eqnarray*}
W(P_{O(n)}):=C_0([0,\infty)
\rightarrow P_{O(n)})
\end{eqnarray*}
 of all continuous mappings 
\begin{eqnarray*}
w:[0,\infty) \rightarrow P_{O(n)},
\end{eqnarray*}
where ${\cal B}(W(P_{O(n)}))$ is the $\sigma$-field on $W(P_{O(n)}))$
generated by the Borel cylinders and ${\cal B}_\tau(W(P_{O(n)}))$ be generated by the Borel cylinders up to time $\tau$. Let $P_r$ be the probability
space law on $W(P_{O(n)})$ of $\tau \rightarrow r(\tau,r,w)$.
Consider for each $w\in W(P_{O(n)})$, the time-dependant $(p,p)$-tensor 
$M(\tau,w)=
(M_{a_1 \ldots a_p}^{b_1\ldots b_p}(\tau,w))$ defined
on $[0,\infty) \times P_{O(n)}$ as
 the solution of the system of equations
\begin{eqnarray*}
{dM_{a_1 \ldots a_p}^{b_1\ldots b_p} \over d\tau}(w(\tau))=   
-{1 \over 2}(-1)^\nu \sum_{\nu = 1}^p {\rm Ric}^{\flat ~b_1}_k(w(\tau))M_{a_1\ldots a_p}^{b_2\ldots b_\nu k b_{\nu + 1}\ldots b_{\nu + p}}(w(\tau))
\end{eqnarray*}
\begin{eqnarray}
-\sum_{1\leq\mu < \nu \leq p}(-1)^{\nu + \mu}\tilde R^{b_2 b_1}_{.i. k}(w(\tau))
M_{a_1 \ldots a_p}^{b_3 \ldots b_{\mu + 1}kb_{\mu + 2}\ldots b_\mu b_{\mu + 1}\ldots b_p}(w(\tau)),
\end{eqnarray}
with
\begin{eqnarray}
M_{a_1 \ldots a_p}^{b_1\ldots b_p}(0) = \delta^{b_1}_{a_1}\ldots \delta^{b_p}_{a_p},\forall 1<a_1<\ldots <a_p \leq n,
\end{eqnarray}
and $\tilde R :=(\tilde R^{ab}_{..cd}) = (R^{\alpha \beta}_{..\nu \mu}
 e^\nu_ce^\mu_d \theta^a_\alpha \theta^b_\beta)$, 
the horizontal lift of the Riemannian curvature $(2,2)$ tensor,
which we shall write as $\tilde R = R\otimes e \otimes e \otimes \theta \otimes \theta$, so that we also have that $R = \tilde R \otimes \theta \otimes \theta \otimes
e \otimes e$, and ${\rm Ric}^\flat = ({\rm Ric}^{\flat}~^b_a)
 = (\tilde R^{bc}_{ca})
\equiv (R^\beta_\alpha e^\alpha_a e^\beta_b)$ which we shall write
 as ${\rm Ric}^\flat = {\rm Ric} \otimes e \otimes \theta$, is the
horizontal lift of the Ricci tensor, to $P_{O(n)}$. For later use, we introduce the tensor 
\begin{eqnarray*}
R^\flat : = (R^{a\beta}_{cd}) , 
\end{eqnarray*}
so that
\begin{eqnarray*}
R^\flat \otimes \theta = \tilde R.
\end{eqnarray*}

It is clear that system $(25,26)$ has a unique solution.
From the uniqueness of the solutions, it follows that
\begin{eqnarray}
M^{\sigma(b_1)\ldots \sigma(b_p)}_{\sigma(a_1)\ldots \sigma(a_p)}(\tau,w) = M^{b_1 \ldots b_p}_{a_1 \ldots a_p}(\tau,w),
\end{eqnarray}
for every permutation $\sigma$.
One can prove still that this solution is $O(n)$-equivariant, since forall $A \in O(n),w\in W(P_{O(n)})$, 
\begin{eqnarray}
M^{b_1\ldots b_p}_{a_1\ldots a_p}(\tau, T_Aw) = M_{c_1\ldots c_p}^{d_1\ldots d_p}(\tau,w)B_{d_1}^{b_1}\ldots B^{b_p}_{d_p}A^{c_1}_{a_1}
\ldots A^{c_p}_{a_p}, 
\end{eqnarray}
where $B = A^{-1}$.
Let now 
\begin{eqnarray*}
\tilde \omega = \tilde \omega (r) ={1 \over p!}\tilde \omega_{b_1\ldots b_p}\theta^{b_1}\wedge \ldots \wedge \theta^{b_p}\in \Lambda^p (T^*P_{O(n)})
\end{eqnarray*}
 be the horizontal lift of a given $\omega =\omega (x) \in \Lambda^p(T^*M)$, and consider the
differential form $\tilde U$, defined on $[0,\infty)\times P_{O(n)}$ with components $\tilde U_{a_1 \ldots a_p}(\tau,r)$ given by
\begin{eqnarray}
\tilde U(\tau,r) := E_r[M(\tau,w(\tau))\otimes \tilde \omega (w(\tau))]
\end{eqnarray}
which componentwise is
\begin{eqnarray}
\tilde U_{a_1 \ldots a_p}(\tau,r) = E_r[M_{a_1 \ldots a_p}^{b_1\ldots b_p}(\tau, w))
\tilde\omega_{b_1\ldots b_p}(w(\tau))].
\end{eqnarray}
This expression is alternate in $(a_1,\ldots,a_p)$ and obviously, it is $O(n)$-equivariant. Consequently, there is a $p$-form, $U$, defined
on $[0,\infty)\times M$, whose horizontal lift is $\tilde U$. If we consider the time dependant $p$-form given by 
\begin{eqnarray*}
M\otimes \tilde \omega := {1 \over p!}
M_{a_1\ldots a_p}^{b_1\ldots b_p}\tilde \omega_{b_1\ldots b_p}\theta^{a_1}\wedge \ldots\wedge \theta^{a_p},
\end{eqnarray*}
and apply to it the Ito formula, we get
\begin{eqnarray*}
(M\otimes \tilde \omega)(\tau,w(\tau))-
(M\otimes\tilde \omega )(0,r)= 
\end{eqnarray*}
\begin{eqnarray}
{\rm a ~martingale~} +
\int^\tau_0 M\otimes [{\partial \tilde \omega \over \partial \tau} + \tilde H_p(g,Q)\tilde \omega )](s,w(s))ds.
\end{eqnarray}
Therefore, 
\begin{eqnarray*}
\tilde U= \tilde U(\tau,r) = {1 \over p!}\tilde U_{a_1\ldots a_p}\theta^{a_1}\wedge \ldots \wedge \theta^{a_p}(\tau,r)
\end{eqnarray*}
 with
components given by
$(30)$ is the unique smooth solution of the Cauchy problem
\begin{eqnarray}
{\partial \tilde U \over \partial \tau} = \tilde H_p(g,Q)\tilde U,
\end{eqnarray}
with initial condition
\begin{eqnarray}
\tilde U(0,r) = \tilde \omega (r).
\end{eqnarray}

{\bf Theorem 1.} The $p$-form $U \in \Lambda^p([0,\infty) \times T^*M)$, such that its horizontal lift coincides
with $\tilde U$, i.e.
$\tilde U \equiv (U(D\pi)) = U\otimes^p e $, where $D$ denotes the derivative and $\otimes^p e$ denotes the $p$-th tensor product with $e$,
is the unique smooth solution of the Cauchy
problem
\begin{eqnarray}
{\partial u \over \partial \tau} = H_p(g,Q)u, u(0,-) =\omega.
\end{eqnarray}

This theorem extends a construction originally due to P.Malliavin, for the Riemannian connection
driftless case [1,7,30]. The process $M$ constructed above, to give account of the Riemannian curvature terms, is called a multiplicative operator functional. 

To conclude with these definitions, we note that from $(6,10 \& 12)$ follows an expression -which we shall use later- for the laplacian on $2$-forms $\phi (r) = \phi_{ab}\theta^a\wedge \theta ^b$
defined on $P_{O(n)}$:
\begin{eqnarray}
\tilde H_p(g,Q)\phi_{ab} = {1\over 2}(L^{\nabla})^2\phi_{ab} + {1\over 2}({\rm Ric}^\flat \phi)_{ab} - (\tilde R \phi)_{ab}.
\end{eqnarray}

\section{The heat equation for manifolds with boundary}
\hspace{2em}
Now, we shall assume further conditions on $M$, that of having a smooth boundary $\partial M$. Near the boundary,
we can choose a coordinate open
neighborhood $U$ of $M$, and smooth coordinate
functions $(x^1,\ldots,x^n)$, such that $x^n \geq 0$, for all $x \in U$; furthermore $x \in U \cap \partial M
$ if and only if $x^n = 0$, and  we shall further assume that we have a smooth Riemannian metric $g=(g_{\alpha\beta})$ on $T^*M$ such that
$g_{\alpha n}(x) =0$, for $\alpha =1,\ldots,n-1$.

Let $\phi$ be a $p$-form on $M$, i.e. $\phi \in \Lambda^p(T^*M)$. Denote by $\phi_{{\rm tan}} = \phi |_{\partial M}$, the tangent component of $\phi$.
and the normal component by $\phi_{{\rm norm}} = \phi - \phi_{{\rm tan}}$. We can still represent the normal component
using duality. Indeed, if $*:\Lambda^p (T^*M) \rightarrow \Lambda^{n-p}(T^*M)$ is the star operator defined by the metric $g$, then
\begin{eqnarray}
(-1)^{p(n-p) + p -1}[*(*\phi \wedge \eta) \wedge \eta]|_{\partial M} = \phi_{{\rm norm}},
\end{eqnarray}
where $\eta (x)$ is a $1$-form such that $\eta|_{\rm \partial M}=g_{\alpha \beta}(x)n^\beta(x)dx^\alpha, x \in \partial M$, where
\begin{eqnarray}
n(x):= n^\alpha (x){\partial \over \partial x^\alpha}= {g^{\alpha n} \over g^{nn}(x)^{1\over 2}}{\partial \over \partial x^\alpha}, \alpha = 1,\ldots ,n, x \in \partial M,
\end{eqnarray}
is the inward pointing normal unit vector to $\partial M$ at $x$.

Then, we shall say that a $p$-form $\phi$ is said to satisfy
the absolute boundary condition if
\begin{eqnarray}
\phi_{{\rm norm}} = 0, (d\phi)_{{\rm norm}}= 0.
\end{eqnarray} 
Note that in the case of a $2$-form on $M$, $\Omega = {1\over 2}\Omega_{\alpha \beta}dx^\alpha \wedge dx^\beta$, the absolute
boundary conditions for $\Omega$ are
\begin{eqnarray}
\Omega_{\alpha n} = 0, \alpha = 1,\ldots,n-1,
\end{eqnarray}
and
\begin{eqnarray} 
{\partial \Omega_{\alpha \beta}\over \partial x^ n} = 0,\alpha,\beta =1,\ldots,n-1,
\end{eqnarray}
respectively.

We wish to solve the boundary-initial-value problem for 
\begin{eqnarray*}
\tilde \Omega ={1\over 2}\tilde \Omega_{a_1a_2}(\tau,r)
\theta^{a_1}\wedge
\theta_{a_2}\in \Lambda^2([0,\infty)\times T^*P_{O(n)})
\end{eqnarray*}
given by
\begin{eqnarray}
{\partial \tilde \Omega \over \partial \tau}(\tau,r) = \tilde H_p(g,Q)\tilde \Omega, ~~~~\tilde \Omega(0,r) = \tilde \omega(r), r\in P_{O(n)},
\end{eqnarray}
where $\tilde \omega$ is the horizontal lift to $P_{O(n)}$ of  
a given $p$-form, $\omega$ defined on $M$, with the absolute boundary conditions. 
In fact, conditions $(39 \& 40)$ can be lifted to $[0,\infty)\times P_{O(n)}$
to give:
\begin{eqnarray}
\theta^{a_2}_\alpha (x)\theta^{a_1}_n (x)\tilde \Omega_{a_1a_2}(\tau,r)|_{\partial P_{O(n)}} = 0, \theta = e^{-1}= (\theta^a _\alpha (x)dx^\alpha),
\end{eqnarray}
\begin{eqnarray}
\theta^{a_1}_{\alpha_1}(x)\theta^{a_2}_{\alpha_2}(x) {\partial \over \partial x^n}\tilde \Omega_{a_1a_2}(\tau,r)|_{\partial P_{O(n)}} =0, a=1,\ldots,n-1.
\end{eqnarray}
where $\partial P_{O(n)} =\{r=(x,e): x \in \partial M\}$.
We shall solve problem $(41,42 \&43)$ using the generalized horizontal Brownian motion on $P_{O(n)}$ and a multiplicative horizontal operator functional.  We start by reformulating the boundary conditions.
 Let $P = (P^\alpha_\beta)$ defined by $P^n_n =1$, and $P^\alpha_\beta = 0$, 
for $\alpha,\beta \ne n$; consider further $Q: = I-P$.  Thus, the absolute boundary conditions $(42 \& 43)$ can be rewritten as the single equation
\begin{eqnarray}
P\otimes \theta\otimes \theta \otimes \tilde \Omega (\tau,r)+ Q\otimes\theta\otimes \theta \otimes
{\partial \tilde \Omega \over \partial x^n}(\tau,r) = 0, \break r= (x,e) \in \partial P_{O(n)}, \theta = e^{-1}.
\end{eqnarray}
Furthermore, we can rewrite the boundary conditions $(44)$, instead of as a condition on $\Lambda^2([0,\infty)\times T^*P_{O(n)})$, as a condition for
\begin{eqnarray*}
\tilde \Omega^\flat (\tau,r) = \tilde \Omega^\flat_\tau(r) = 
{1 \over 2}\tilde \Omega^\flat _{\alpha a}(\tau,r)dx^\alpha \wedge \theta^a (x) \in L(TM, R^{n*}), \forall \tau \geq 0, 
\end{eqnarray*}
i.e. for each $\tau \ge 0$ we have a linear mapping from
$TM \rightarrow R^{n*}$ such that for each $r=(x,e) \in P_{O(n)}$, $\tilde \Omega^\flat _\tau(r)$
maps $T_xM$ into ${\rm hor}~(T^*P_{O(n)})\simeq R^{n^*}$, related to $\tilde \Omega (\tau,r)$ by 
\begin{eqnarray*}
e \otimes \tilde \Omega^\flat \equiv \tilde \Omega
\end{eqnarray*}
and satisfying the
boundary conditions
\begin{eqnarray*}
P\otimes \theta \otimes \tilde \Omega^\flat _\tau(r) + Q\otimes \theta \otimes [{\partial \tilde \Omega^\flat_\tau \over \partial x^n}(r) -
 {\partial \tilde \Omega^\flat _\tau \over \partial e^\alpha_m}(r)  \Gamma^\alpha _{n\beta}(x) e^\beta_m(x)
\end{eqnarray*}
\begin{eqnarray}
 +
 e(x)\otimes \Gamma_n (x)\otimes \theta (x)\otimes \tilde \Omega^\flat _\tau](r) = 0, \forall r =(x,e) \in \partial P_{O(n)},
\end{eqnarray} 
where
$\Gamma_n (x) = (\Gamma^\alpha_{n\beta}(x))$,  and $(\theta \otimes \tilde \Omega^\flat)_{\alpha \beta}  
= ({1\over 2}\theta ^a_\beta \tilde \Omega^\flat_{\alpha a})$
is the multiplication in the representation space $R^{n^*}$ of this linear mapping.
Indeed, note that if $\tilde \Omega ^\flat (\tau,r)$ is $O(n)$-equivariant, then
 $\tilde \Omega^\flat  (\tau,r) =   \Omega(\tau,x) \otimes e(x), r=(x,e) \in P_{O(n)}$ \footnote{This identity, which can be rewritten as
$\Omega(\tau,x) = \tilde\Omega^\flat (\tau,r)\otimes \theta (x)$, or still $\tilde \Omega (\tau,r) = \Omega (\tau,x)\otimes e \otimes e$,
will play a crucial role for obtaining
the expression for the vorticity as defined on $M$ instead of $P_{O(n))}$; we shall work out the boundary-initial-value problem
for Navier-Stokes on $[0,\infty) \times L(TM, T^*P_{O(n)})$, to later project on $\Lambda^ 2 ([0,\infty)\times T^*M)$ by this procedure.}, and
 thus
\begin{eqnarray}
{\partial \tilde \Omega^\flat_\tau \over \partial e^\alpha_m}  \Gamma^\alpha_{n\beta}(x)e^\beta_m(x)= e \otimes\Gamma_n(x) \otimes
\theta \otimes \tilde \Omega^\flat
_\tau.
\end{eqnarray}
Therefore , in account of identity $(46)$, by further multiplying $(45)$ by $\theta$, we retrieve the boundary conditions $(44)$
for equivariant $2$-forms in $\Lambda^2 ([0,\infty)\times T^*P_{O(n)})$, or multiply by $\theta$ on the right, to obtain a $2$-form in
$\Lambda^2 ([0,\infty)\times T^*M)$ as explained in the previous footnote.

From a localization argument, we can assume for simplicity
that $M$ is the upper half-space of $R^n$, $\{x:x=(x^1,\ldots,x^n)\in R^n\}$
and $\partial M = \{x \in M: x^n = 0\}$. The metric $g$ on $M$ is assumed further to be bounded
and smooth together with all its derivatives, i.e. $g_{\alpha \beta} \in C^\infty_b (M).$, $1\leq \alpha \leq \beta \leq n$.

We start by considering the following Stratonovich stochastic differential equation for the diffusion
process $r(\tau) = (X(\tau),e(\tau))$ on $P_{O(n)}\sim R^n_+\times R^{n^2}$, following [1]:
\begin{eqnarray}
dr^\alpha (\tau) = L^\nabla_k(r(\tau))\circ dB^k(\tau) + \delta ^{\alpha}_nd\phi (\tau), r(0) = r, \alpha =1,\ldots,n,
\end{eqnarray}
where as before, $\nabla$ denotes a Riemann-Cartan-Weyl connection with coefficients of the form $(1)$, $L^\nabla_k ,k=1,\ldots,n$
denote the canonical horizontal vectorfields defined by this connection on $P_{O(n)}$ in $(4)$, and $B$ denotes a $n$-dimensional
Brownian motion. We note that it can 
be rewritten,  for $r=(X,e)$, as the system
\begin{eqnarray}
dX^\alpha (\tau) = e^\alpha_k (\tau)\circ dB^k(\tau) + \delta ^{\alpha}_nd\phi (\tau), X(0) = x, \alpha = 1,\ldots,n
\end{eqnarray}
\begin{eqnarray}
de^\alpha_ k(\tau) = &
-& \Gamma^\alpha _{\beta \gamma}(X(\tau))e^\gamma_k(\tau)\circ dX^\beta(\tau) =\nonumber\\
& -& \Gamma^\alpha _{\beta \gamma}(X(\tau))e^\gamma_k(\tau)e^\beta_c(\tau)\circ dB^c(\tau)\nonumber\\ 
&-& \Gamma^\alpha _{n \gamma}(X(\tau))e^\gamma_k(\tau)d\phi (\tau), e^\alpha (0)= e(x) \in T_xM, \alpha, k=1,\ldots,n.
\end{eqnarray}
Here, $\phi(\tau)$ is a continuous non-decreasing process which increases only when $X(\tau) \in \partial M$, i.e. a local time on
$\partial M$, and which causes the reflection of the process at the boundary.

This is a generalized (i.e. with non-zero drift described by $\tilde (Q^g)$, the horizontal lift of $Q^g$) horizontal Brownian motion 
on the bundle of orthonormal frames $P_{O(n)}$ with reflecting boundary. Indeed, it can be easily seen that if $(X(0),r(0)) \in P_{O(n)}$, then
$(X(\tau),r(\tau)) \in P_{O(n)}$. We readily note that if $\partial M = 0$ and if we further set
$\phi=0$ in eq. $(47)$, we obtain the generalized horizontal Brownian motion on $P_{O(n)}$, for a manifold without boundary. Having assumed that 
$g_{\alpha \beta}\in C^\infty_b (M)$, if we assume further that
$Q_\alpha \in C^\infty_b (M)$ ($1 \le \alpha \le n)$ and $\phi \in C^2_b(M)$, then eqt. $(47)$ has a unique strong solution (c.f. Theorem $7.2$, chapter IV, [1]).

Set now
\begin{eqnarray}
A^{nn}(r) = g^{n n}(x), A^{n,\alpha}_k(r) = -e^\beta_k (x)
\Gamma^\alpha_{n \beta}(x)g^{nn}(x), r=(x,e).
\end{eqnarray}
The following result is crucial to our following constructions.

{\bf Theorem 2.}Let $r(\tau) = (X(\tau),e(\tau))$, the generalized horizontal Brownian motion with reflecting boundary condition described by eqt. $(47)$.
 For any smooth function $F(\tau,r)$ defined on $[0,\infty)\times P_{O(n)}$ we have
\begin{eqnarray*}
dF(\tau,r(\tau)) = [(\tilde H_0(g,Q)F)(\tau,r(\tau)) + {\partial F \over \partial \tau}(\tau,r(\tau))]d\tau
\end{eqnarray*}
\begin{eqnarray}
+ (L^\nabla_k F)(\tau,r(\tau))dB^k(\tau) + (\tilde X_n F)(\tau,r(\tau))d\phi (\tau),
\end{eqnarray}
where $\tilde X_n$ is the canonical horizontal lift of $X_n= {\partial \over \partial x^ n}$, i.e.
\begin{eqnarray}
(\tilde X_n F)(\tau,r) = {\partial F \over \partial x^n}(\tau,r) +
{A^{n,\alpha}_k(r) \over A^{nn}(r)}{\partial F \over \partial e^\alpha _k}(\tau,r).
\end{eqnarray}

{\bf Proof.} It follows from the Ito formula that
\begin{eqnarray*}
dF(\tau,r(\tau)) & = & e^\alpha_a(\tau){\partial F \over \partial x^\alpha}(\tau,r(\tau))\circ dB^a(\tau)\nonumber\\
 & - &
{\partial F \over \partial e^\alpha _j}(\tau,r(\tau))\Gamma^\alpha _{\beta \nu}(X(\tau))e^\nu_j(\tau)e^\beta_a(\tau)\circ dB^a(\tau)
\end{eqnarray*}
\begin{eqnarray*}
&  +&  {\partial F \over \partial \tau}(\tau,r(\tau))d\tau + {\partial F \over \partial x^n}(\tau,r(\tau))d\phi(\tau)\nonumber\\
&  - &   {\partial F \over \partial e^\alpha_a}(\tau,r(\tau))\Gamma^\alpha_{n\kappa}(r(\tau))e^\kappa_a d\phi(\tau)\nonumber\\
& = & (L_a^\nabla F)(\tau,r(\tau))\circ dB^a(\tau) + {\partial F \over \partial \tau}(\tau,r(\tau))d\tau + \tilde X_n(F)(\tau,r(\tau))d\phi(\tau)
\end{eqnarray*}
so that 
\begin{eqnarray*}
d(L^\nabla_a F)(\tau,r(\tau))  & = & L^\nabla_b(L^\nabla_a F)(\tau,r(\tau))\circ dB^b(\tau)\nonumber\\
& + & {\partial (L^\nabla_a F) \over \partial \tau}(\tau,r(\tau))d\tau + \tilde X_n(L^\nabla_a F)(\tau,r(\tau))d\phi(\tau)
\end{eqnarray*}
and then
\begin{eqnarray*}
(L^\nabla _a   F)(\tau,r(\tau)\circ dB(\tau)  = 
 + {1\over 2}d(L^\nabla_a F)(\tau,r(\tau))\bullet dB^a(\tau)+ (L^\nabla_a F))\tau,r(\tau)) dB^a(\tau)
\end{eqnarray*}
\begin{eqnarray*}
& = & {1\over 2}(L^\nabla_a )^2F(\tau,r(\tau))d\tau
+  (L^\nabla_a F)(\tau,r(\tau))\bullet dB^a(\tau)\nonumber\\
& = &{1 \over 2}\tilde H_0(g,Q)F(\tau,r(\tau)) d\tau
 + (L^\nabla_a F)(\tau,r(\tau)) dB^a(\tau),
\end{eqnarray*}
the last identity follows from $(6)$, and $\bullet$ denotes the Ito contraction on stochastic differentials [1], with which we can conclude with the proof.

From Theorem $2$ follows that the process $r(\tau) \in P_{O(n)}$ is determined by the differential generator $\tilde H_0(g,Q)$ with the boundary condition
$\tilde X_n F = 0$ on $\partial P_{O(n)}$. Furthermore, 
\begin{eqnarray}
dX^ n(\tau)\bullet dX^n(\tau) = A^{nn}(r(\tau))d\tau, 
\end{eqnarray}
\begin{eqnarray}
dX^n(\tau)\bullet de^\alpha_k(\tau) = A^{n,\alpha}_k(r(\tau))d\tau, \alpha,k = 1,\ldots,n.
\end{eqnarray}

To obtain a solution of the boundary-initial-value problem $(41 \& 45)$ for the case of smooth boundary, we need to construct a multiplicative operator functional, which extends the one presented for
the boundaryless case presented in Theorem $1$. Consider a canonical realization of the generalized horizontal Brownian motion on $P_{O(n)}$ with reflecting boundary
condition. We thus consider $W(P_{O(n)}) = C([0,\infty)\rightarrow P_{O(n)})$, provided with the probability measure $P_r$ of the solution of
$(46)$ and, $r(\tau,w) = w(\tau), \forall w \in W(P_{O(n)})$. Given a Borel probability measure $\mu$ on $P_{O(n)}$, we define
$P_\mu(B) = \int_{P_{O(n)}} P_r(B)\mu(dr), \forall B \in {\cal B}(W(P_{O(n)}))$. Let ${\cal F} = \cap_{\mu}{\cal B}(W(P_{O(n)}))^{P_\mu}$ and ${\cal F}_\mu
=\{A \in {\it F}$: for any $ \mu$, there exists  $B_\mu \in {\cal B}_\tau (W(P_{O(n)}))$, such that $P_\mu (A\triangle B)=0\}$. In the following we fix
$\mu$ and consider the probability space $(W(P_{O(n)}),{\cal F},P_\mu)$. Writing for simplicity $r(\tau)\equiv
r(\tau,w) = (X(\tau,w),e(\tau,w)) \equiv (X(\tau),e(\tau))$, we set
\begin{eqnarray}
\phi(\tau) = lim_{\epsilon \rightarrow 0}{1\over 2\epsilon}\int^ \tau_0 I_{[0,\epsilon)}(X^n(s))g^{nn}(x)ds,
\end{eqnarray}
and
\begin{eqnarray}
B^i(\tau) = \int^\tau_0 \theta^i_\alpha (s)\circ [dX^\alpha (s) - \delta^\alpha_n d\phi(s)],
\end{eqnarray}
where $\theta = (\theta^a _\alpha (x)dx^\alpha)$ is the dual (co-tetrad) field to $e$.
Then, $\{B^i(\tau): i=1,\ldots,n\}$ is an $n$-dimensional $({\cal F}_\tau)$-Brownian motion and $(\{r(\tau) = r(\tau,w)\},\phi(\tau))$
satisfies the s.d.e. equation $(47)$.

{\bf Lemma 1.} $\{P_r\}$ is invariant under the action of $O(n)$, i.e. if we define
\begin{eqnarray}
(T_A w)(\tau) = T_A(w(\tau)),\forall w \in w(P_{O(n)}),\tau \geq 0
\end{eqnarray}
and if $T_A(P_r)$ is the image of $P_r$ under the mapping $w \rightarrow T_Aw$, then,
\begin{eqnarray}
T_a(P_r) = P_{T_A(r)}.
\end{eqnarray}

{\bf Proof:}
Let $r(\tau)$ be a solution of eq. $(47)$ with $B(\tau)$ and $\phi (\tau)$ such that $r(0) = r$. Then 
for $A \in O(n)$, $\tilde r(\tau) :=
T_A(r(\tau))$ is a solution of eq. $(47)$ with $\tilde B(\tau) = A^{-1}B(\tau)$ and $\tilde \phi (\tau) = \phi (\tau)$, 
such that $\tilde r(0)= Ar$. If we further observe
that $\tilde B(\tau)$ is another $n$-dimensional Brownian motion, 
then $T_A(P_r) = P_{T_A(r)}$ holds by the uniquenesss of solutions.

As a consequence of Lemma $1$, we note that $X(\tau) = \pi (r(\tau))$ defines a diffusion process on $M$ whose differential generator is $H_0(g,Q)$, and with boundary
condition given by the vanishing of the normal derivative of $f$ at $x \in \partial M$:
\begin{eqnarray}
{\partial f \over \partial n}  := n^\alpha (x){\partial f\over \partial x^\alpha} = 0, x \in \partial M.
\end{eqnarray}
This diffusion is the so-called generalized Brownian motion on $M$ with reflecting boundary condition; it represents the random process on actual configuration space, accounting with
the given boundary conditions. 

 In the following we
shall fix a probability measure $\mu$ and consider the probability space $(W(P_{O(n)}),{\cal F},P_\mu)$. 
Then $r(\tau) =r(\tau,w) = (X(\tau,w),e(\tau,w))\equiv (X(\tau),e(\tau))$
is the solution of eq. $(47
)$ with $B(\tau)$ and $\phi(\tau)$ given by $(55 \& 56)$. 

We shall consider the $R^n\otimes R^{n^*}$-valued
process $K(\tau) = (K^{\alpha}_\gamma(\tau,w))$ defined -in extending a construction due to H. Airault (see [3])- 
as follows:
Set $K^1(\tau) = K(\tau)\otimes P$ and $K^2(\tau) = K(\tau)\otimes Q$; hence $K(\tau) = K^1(\tau) + K^2(\tau)$, where  
\begin{eqnarray*}
(i) {\rm for ~any~} \tau \geq 0 ~{\rm such ~that~} X(\tau) \in {\rm int}(M),
\end{eqnarray*}
\begin{eqnarray}
dK^1(\tau) :& = & dK(\tau)\otimes P = K(\tau)\otimes [ \theta(\tau)\otimes de(\tau)\nonumber\\
& +
& \{{1\over 2}{\rm Ric}(X(\tau)) - R(X(\tau))\}d\tau ] \otimes P,
\end{eqnarray}
\begin{eqnarray}
(ii) {\rm for~ any~} \tau \geq 0 &,  & dK^2(\tau): = dK^2(\tau)Q = K(\tau)\otimes [\theta(\tau)\otimes de(\tau)\nonumber\\
& + &\{{1\over 2}{\rm Ric}(X(\tau)) -  R(X(\tau))\}d\tau ] I_{{\rm int(M)}}(X(\tau))\otimes Q,
\end{eqnarray}
and with probability $1$, $\tau \rightarrow K^1(\tau)$ is right-continuous with left-hand limits; furthermore
\begin{eqnarray}
(iii)~ K^1(\tau) = 0, ~{\rm if~} X(\tau) \in \partial M,
\end{eqnarray}
and the initial value is
\begin{eqnarray}
(iv)~  K^1(0) = I_{{\rm int (M)}}(X(0))e(0)\otimes P, K^2(0) = e(0)\otimes Q.
\end{eqnarray}
Note that in terms of a component representation for $K$, we have that the above expressions appear to be of the form
\begin{eqnarray*}
(dK)^{1\gamma}_\alpha P_\gamma ^\delta &=& K_\alpha ^\beta [(\theta \otimes de)_\beta^\gamma P_\gamma^\delta
 + (R _\beta^\gamma + R^{\gamma \epsilon}_{\beta \kappa})P^\kappa_\delta d\tau\nonumber\\
& = & K_\alpha^\beta [(\theta \otimes de)^\gamma_\beta P_\gamma^\delta
+ (\theta ^b_\beta R^a_b e^\gamma_a +  \theta ^c_\beta R^{a \epsilon}_{c \kappa}e^\gamma_a)P^\kappa_\delta d\tau].
\end{eqnarray*}
and similarly for $K^2$. ~~\footnote{It may seem that instead of taking $R$ we should take the tensor with components
$(R^{\alpha \beta}_{\gamma d}) = (R^{\alpha \beta}_{\gamma \delta}e^\delta_d)$, yet on multiplying 
with $P$ and $Q$ it turns to be indistinct with our previous choice.}

An equivalent formulation is the following. An $R^n \otimes R^{n^*}$-valued process adapted to $({\cal F}_\tau)$ is a solution
 of the above s.d.e.
with the given initial conditions iff
\begin{eqnarray*}
K^1(\tau) &= & K(\tau)\otimes P = I_{\{\tau <\sigma \}}(e(0) + \int^\tau_0 K(s)\otimes[\theta (s)\otimes de(s)\nonumber\\
 &+& \{{1\over 2}{\rm Ric}(X(s)) -  R(X(s))\}ds])\otimes P
\end{eqnarray*}
\begin{eqnarray}
+ I_{\{\tau \geq \sigma\}}\int_{t(\tau)}^\tau K(s)\otimes [\theta (s)\otimes de(s) +\{{1\over 2}{\rm Ric}(X(s)) -  R(X(s))\}ds]\otimes P,
\end{eqnarray}
\begin{eqnarray*}
K^2(\tau) = K(\tau)\otimes Q = e(0)\otimes Q + \int_0 ^ \tau K(s)\otimes [\theta (s)\otimes de(s)
\end{eqnarray*}
\begin{eqnarray}
+\{{1\over 2}{\rm Ric}(X(s)) -  R(X(s))\} ds ]I_{{\rm int(M)}}(X(s))\otimes Q,
\end{eqnarray}
where
\begin{eqnarray}
\sigma = {\rm inf}\{s:X(s) \in \partial M\}, 
\end{eqnarray}
 and equal to $0$ if this set is empty,
is the first-hitting time of $\{X(\tau):\tau \ge 0\}$ to $\partial M$, and
\begin{eqnarray}
t(\tau) = {\rm sup}\{s:s\leq \tau, X(s) \in \partial M\},
\end{eqnarray}
and equal to 0  if this set is empty,
is the last-exit time from $\partial M$.

Let $\Xi$ be the set of all $R^n \otimes R^{n^*}$-valued processes $\xi(\tau)$ defined 
on $(W(P_{O(n)}),\break {\cal F},P)$
adapted to $({\cal F}_\tau)$ such that $\tau \rightarrow \xi (\tau)$ is right continuous with left-hand limits almost surely and satisfies
\begin{eqnarray}
{\rm sup}_{\tau \in [0,T]}E_\mu[||\xi(\tau)||^2] < \infty, \forall T>0.
\end{eqnarray}
Define a mapping $\Phi :\Xi \rightarrow \Xi$
\begin{eqnarray*}
\Phi^1(\xi)(\tau):& =& \Phi(\xi)(\tau)P =I_{\{\tau <\sigma \}}(e(0) + \int^\tau_0 K(s)\otimes [\theta (s)\otimes de(s)\nonumber\\
& +& ({1\over 2}{\rm Ric}(X(s)) -  R(X(s)))ds])\otimes P
\end{eqnarray*}
\begin{eqnarray}
+ I_{\{\tau \geq \sigma\}}\int_{t(\tau)}^\tau K(s)\otimes [\theta (s)\otimes de(s) +({1\over 2}{\rm Ric}(X(s)) - R(X(s)))ds]\otimes P,
\end{eqnarray}
\begin{eqnarray}
\Phi^2(\xi)(\tau):& = & \Phi(\xi)(\tau)Q=e(0)Q + \int_0^\tau K(s)\otimes [\theta (s)\otimes de(s)\nonumber\\
& + &({1\over 2}{\rm Ric}(X(s)) -  R(X(s)))ds]I_{{\rm int(M)}}(X(s))\otimes Q.
\end{eqnarray}

Let $A(\tau)$ be the right-continuous inverse of $\tau \rightarrow \phi(\tau)$; set $D=\{s: 0\leq s,
A(s-)<A(s)\}$. If $\tau >0$ is fixed, then
$t(\tau) =A(\phi(\tau)-)$ a.s. Now for $g(\tau)$ an $({\cal F}_\tau)$-well measurable process 
such that
$\tau \rightarrow E_\mu[g(\tau)^2]$ is locally bounded then (see Theorem $(6.6)$ in Ikeda and Watanabe [1])
\begin{eqnarray*}
E_\mu[\{\int^\tau_{t(\tau)}g(s)dB^k(s)\}^2]
\end{eqnarray*} 
\begin{eqnarray*}
\leq  E_\mu[\sum_{u \in D}\{\int^{A(u)\wedge s}_{A(u-)\wedge s}g(s)dB^k(s)\}^2]
=  E_\mu[\int^\tau_0 g(s)^2ds].
\end{eqnarray*}
Then, for every $T>0$ there is a constant $K=K(T)$ such that
\begin{eqnarray}
E_\mu[||\Phi(\xi)(\tau)||^2] \leq K(1 + \int_0^\tau E_\mu[||\xi(s)||^2]ds), \forall \tau \in [0,T].
\end{eqnarray}
This proves that if $\xi \in \Xi$, then $\Phi(\xi) \in \Xi$. Again, using Theorem 6.6 of [1], we have for $\xi_1 ,\xi_2 \in \Xi$, that
\begin{eqnarray}
E_\mu [||\Phi(\xi_1)(\tau) - \Phi(\xi_2)(\tau)||^2] \leq k \int^\tau_0E_\mu [||\xi_1 (s) - \xi_2(s)||^2]ds, \tau \in [0,T].
\end{eqnarray}
Then, from standard arguments follows the following result.

{\bf Theorem 3.} The stochastic differential equation $(60-63)$ has a solution $K(\tau) \in \Xi$ which is unique.

Let $K(\tau)$ be the solution of $(60-63)$; define
\begin{eqnarray}
M(\tau,w) : = K(\tau,w)\otimes \theta(\tau,w) = (K(\tau,w)_\gamma^{\beta}\theta_\beta ^b(\tau,w)), \tau \geq 0, 
\end{eqnarray}
where we recall $\theta$ is the co-tetrad field dual to $e$. 

{\bf Theorem 4. } $M(\tau,w)$ is an $R^n\otimes R^{n^*}\simeq L(TM,R^{n^*})$
-valued multiplicative horizontal functional of the
generalized horizontal Brownian motion on $P_{O(n)}$ with reflecting boundary, i.e.
\begin{eqnarray}
(i) M(\tau,w) ~{\rm is}~ ({\cal F}_\tau)-{\rm ~adapted},
\end{eqnarray}
\begin{eqnarray}
(ii) \forall \tau,s \geq 0, M(\tau + s,w) = M(s,w)M(\tau,\alpha_sw), ~{\rm almost ~surely}
\end{eqnarray}
where the shift operator $\alpha_s:W(P_{O(n)})
 \rightarrow W(P_{O(n)})$ is defined by $(\alpha_s w)(\tau) = w(\tau + s)$

{\bf Proof:} Part $(i)$ is obvious by construction. We now fix $s$ and consider $\tilde K(\tau):=
K(\tau + s,w), \tilde X(\tau): = X(\tau+s,w)$ and $\tilde e(\tau):= e(\tau+s,w)$. Thus, $\tilde K(\tau)$
satisfies eqts. $(60-63)$ with respect to $(\tilde X(\tau),\tilde e(\tau))$. Still, if we apply the shift operator
$\alpha_s$ to $K(\tau)$ we obtain that $\kappa (\tau) = K(\tau,\alpha_sw)$ satisfies eqts. $(60-63)$
with respect to $(X(\tau,\alpha_sw),e(\tau,\alpha_sw)) = (\tilde X(\tau),\tilde e(\tau))$. If we set
\begin{eqnarray*}
K'(\tau) : = K(s,w)\otimes \theta(s,w)\otimes\kappa (\tau),
\end{eqnarray*}
then,
\begin{eqnarray*}
K'(0) &=& K(s,w)\otimes \theta(s,w)\otimes (I_{{\rm int (M)}}(X(s,w))\otimes e(s,w) \otimes P + e(s,w)\otimes Q)\nonumber\\
& = & K(s,w)\otimes (I_{{\rm int(M)}}(X(s,w))\otimes P + Q) =K(s,w),
\end{eqnarray*}
by $(63)$ and the fact that $\theta \otimes e \equiv I$, by definition. Hence $\tilde K(\tau)$ and $K'(\tau)$ satisfy both eqts. 
$(60-63)$ and consequently, they coincide for every $\tau$.
Then,
\begin{eqnarray*}
K(\tau+s,w) = K(s,w)\otimes \theta (s,w)\otimes K(\tau,\alpha_sw),
\end{eqnarray*}
and thus multiplying by $\theta (\tau+s,w) = \theta (\tau,\alpha_sw)$ on the right yields the second part of this Theorem. c.q.d.

An important property of $M = (M(\tau,w))$ is the following Lemma.

{\bf Lemma 2.} If $X(0) \in \partial M$, then
\begin{eqnarray}
P\otimes \theta(0)\otimes M(\tau) = 0, \forall \tau \geq 0.
\end{eqnarray}

{\bf Proof:} It is enough to prove that $P\otimes \theta (0)\otimes K(\tau) = 0, \forall \tau \geq 0$. If $X(0) \in \partial M$, then
\begin{eqnarray}
P\otimes \theta(0)\otimes K(0) & = & P\otimes \theta(0)(I_{{\rm int}(M)}(X(0))e(0)\otimes P + e(0)\otimes Q) \nonumber\\
&=& P\otimes Q = PQ= 0.
\end{eqnarray}
Since $\tilde K(\tau) = P\otimes \theta(0)\otimes K(\tau)$ satisfies eqts. $(60-63)$ with respect to
$\tilde X(\tau),\tilde e(\tau))$, then $\tilde K(\tau) = 0$ by the uniqueness of solutions. 

{\bf Lemma  3.} $M(\tau,T_Aw) = AM(\tau,w)A^{-1}, \tau \geq 0, A \in O(n)$.

{\bf Proof:} Since $X(\tau,T_Aw) = X(\tau,w)$ and $e(\tau,T_Aw) = Ae(\tau,w)$, it follows from the uniqueness of the 
solutions of eqts. $(60-63)$, that $K(\tau,T_Aw) = AK(\tau,w)$. Thus, $M(\tau,T_Aw) = AK(\tau,w)\otimes \theta(\tau,w)A^{-1}
= AM(\tau,w)A^{-1}$.

{\bf Proposition 1.}  For any $\tau \geq 0$ such that $X(\tau) \in {\rm int}(M)$, $M(\tau)= M(\tau,w)$ satisfies
\begin{eqnarray}
dM(\tau) = M(\tau)\otimes [{1 \over 2}{\rm Ric}^\flat (r(\tau) - R^\flat (r(\tau))]d\tau.
\end{eqnarray}

{\bf Proof:}
\begin{eqnarray*}
dM(\tau) = K(\tau)\otimes \theta(\tau)\otimes [de(\tau)\otimes \theta(\tau) + e(\tau)\otimes d\theta(\tau) + de(\tau)\bullet d\theta(\tau)]
\end{eqnarray*}
\begin{eqnarray*}
+  K(\tau)\otimes \theta(\tau)]\otimes [{1 \over 2}{\rm Ric}(r(\tau)) - R(r(\tau))]d\tau
\end{eqnarray*}
\begin{eqnarray*}
=  K(\tau)\otimes \theta(\tau)\otimes d(e(\tau)\otimes \theta(\tau)) + K(\tau)\otimes\theta(\tau)[{1\over 2}{\rm Ric}(r(\tau)) 
-  R(r(\tau))]d\tau
\end{eqnarray*}
\begin{eqnarray*}
=  K(\tau)\otimes \theta(\tau)\otimes [{1 \over 2} {\rm Ric} - R](r(\tau))d\tau 
\end{eqnarray*}
(since $d(e(\tau))\otimes \theta(\tau)) = d({\rm constant}) = 0$)
\begin{eqnarray}
= M(\tau)\otimes [{1\over 2}{\rm Ric}^\flat (r(\tau)) - R^\flat (r(\tau))]d\tau,
\end{eqnarray}
where the last identity follows trivially from the definitions of ${\rm Ric}^\flat$ and $R^\flat$, since componentwise we have the equation for $M = (M_\alpha ^d) = (K_\alpha ^\gamma \theta_\gamma^d)$ 
\begin{eqnarray*}
dM_\alpha^d & = & 
(K^\beta_\alpha\theta^b_\beta R^a_be_a^\gamma \theta^d_\gamma
 +  K^\beta_\alpha \theta^c_\beta R^{a\epsilon}_{c \kappa}e^\gamma_a\theta ^d_\gamma) d\tau \nonumber\\
& = &  
(K^\beta_\alpha \theta ^b_\beta R^a_b\delta_a^d + K^\beta_\alpha \theta^c_\beta R^{a\epsilon}_{c\kappa}\delta^d_a)d\tau\nonumber\\
& = & (M_\alpha ^bR_b^d + M_\alpha ^c R_{c\kappa}^{d\epsilon})d\tau,
\end{eqnarray*}
where we finally note that the r.h.s. of $(78)$ coincides with
 the coupling of $M$ to the Weitzenbock term in which $\tilde R$ has been substituted by $R^\flat$.

Let $C_0(P_{O(n)} \rightarrow L(TM,R^{n^*}))\simeq T^*M \otimes T^*P_{O(n)}$ 
be the set of all bounded continuous functions $F(r)$ on $P_{O(n)}$ taking values in $L(TM,R^{n^*})$, 
and such that 
\begin{eqnarray}
P\otimes \theta \otimes F(r) = 0, ~{\rm if}~ r=
(x,e) \in \partial P_{O(n)}, \theta = e^{-1},
\end{eqnarray}
where again as before we note that  the $\otimes$-product with $F$ is meant in the image, $R^{n^*}$ of the $2$-form
$F(r)={1\over 2}F_{\alpha a}dx^\alpha \wedge \theta ^a(x)$
defined on $TM\wedge R^n$ by $F$.  

For $\tilde \omega \in C_0(P_{O(n)}\rightarrow L(TM,R^{n^*}))$ and $\tau \geq 0$, set
\begin{eqnarray}
(H_\tau \tilde \omega )(r) := E_r[M(\tau,w)\otimes \tilde \omega (r(r,\tau,w))].
\end{eqnarray}
Componentwise, we have that
\begin{eqnarray*}
(H_\tau \tilde \omega )_{\alpha \beta}(r) := E_r[M(\tau,w)_\alpha ^a \tilde \omega _{\beta a}(r(r,\tau,w))].
\end{eqnarray*}

{\bf Theorem 5.} $\{H_r\}$ defines a one-parameter semigroup of operators on $C_0(P_{O(n)}\rightarrow L(TM,R^{n^*}))$.

{\bf Proof:} From Lemma $(1)$
we obtain that for $ r \in \partial P_{O(n)}$, $P\otimes \theta\otimes (H_\tau \tilde \omega)(r) =0$. 
From the uniqueness of the solutions of the s.d.e. $(60-63)$ we deduce the continuity
of the mapping $r \rightarrow P_r \in {\cal F}(W(P_{O(n)}))$, where ${\cal F}(W(P_{O(n)}))$ is the set of all probabilities on $W(P_{O(n)})$ with the
 weak convergence topology;
consequently, the functions $(H_\tau \tilde\omega)(r)$ are continuous in $r$. The equivariance of $\{H_r\}$ follows from Lemmas $1$ and $2$.

We are now in conditions of stating the solution of our boundary-initial-value problem $(41 \& 45)$, extending thus Theorem $1$ 
to the case of smooth boundaries.

{\bf Theorem  6.} Let $F(\tau,r)$ be a $L(TM,R^{n^*})$-valued smooth functions on $[0,\infty) \times P_{O(n)}$ such that for each $\tau \geq 0$, the mapping
$r \rightarrow F(\tau,r)$ is a function in $C_0(P_{O(n)}\rightarrow L(TM,R^{n^*}))$. Then, with probability one, 
\begin{eqnarray*}
M(\tau)\otimes F(\tau,r(\tau))  -  M(0)\otimes F(0,r(0)) = \int^\tau_0 M(s)\otimes (L^\nabla _a F)(s,r(s))dB^a(s)~ +
\end{eqnarray*}
\begin{eqnarray*}
\int^\tau_0 M(s)\otimes [{\partial F \over \partial \tau}(s,r(s)) + \tilde H_0F(s,r(s))
 +  ({1\over 2}{\rm Ric}^\flat (r(s)) -  R^\flat (r(s))\otimes F(s,r(s))]ds
\end{eqnarray*}
\begin{eqnarray*}
+ \int^\tau_0 M(s)\otimes e(s)\otimes Q\otimes \theta(s)\otimes[{\partial F \over \partial x^n}(s,r(s)) -{\partial F \over \partial e^\alpha_m}(s,r(s)) \Gamma^\alpha_{n\beta}(X(s)) e^\beta_m(s)
\end{eqnarray*}
\begin{eqnarray}
+e(s)\otimes \Gamma_n(X(s))\otimes \theta(s)\otimes F(s,r(s))]d\phi(s).
\end{eqnarray}

{\bf Proof.} The proof is an extension of a similar result for $1$-forms, Theorem $6.5$ in [1], and is almost identical: One has to replace the expression
${1\over 2}{\rm Ric}^\flat$ everywhere it appears by ${1\over 2}{\rm Ric}^\flat - R^\flat$.

Note that the r.h.s. term of eq. $(82)$ in which appears the coupling of $F$ to the curvature terms , which consistent with our previous notations can be thought as an object of the form
${1\over 2}\tilde \Omega^\flat _{\alpha a}dx^\alpha \wedge \theta^a$ coupling
with ${1\over 2}{\rm Ric}^\flat -R^\flat$, coincides with the 
coupling of the Weitzenbock potential term
${1\over 2}{\rm Ric}^\flat - \tilde R$ 
to an equivariant $2$-form $\tilde F$  on $P_{O(n)}$ induced by $F$ from the identity $e\otimes F = \tilde F$, 
or with our previous notation a $2$-form $\tilde \Omega$
which satisfies $\tilde \Omega = e \otimes \tilde \Omega ^\flat$: Indeed,
$R^\flat F = \tilde R \tilde F$, and already ${\rm Ric}^\flat \otimes F \equiv {\rm Ric}^\flat \tilde F$, so that this term expresses the multiplicative operator functional for the zero-boundary case, while the last two terms in the identity
$(82)$ give precise account of the boundary conditions. Therefore, Theorem $6$ can be regarded as a martingale problem solution [40] to the boundary-initial value problem
$(41 \& 45)$;  thus, we have proved that $H_\tau \tilde \omega$ is the solution to this problem.

\section{The Navier-Stokes equations for the vorticity and the absolute boundary conditions}
\hspace{2em}
We consider an oriented smooth connected manifold $M$ with smooth boundary, $\partial M$, provided with a Riemannian metric, $g$,
and with a time-dependant $1$-form
$u(\tau,x) = u_\tau(x)$, $\tau \ge 0$, $x \in M$.
The Navier-Stokes equations for the velocity time-dependant $1$-form $u_\tau (x)$ satisfying the incompressibility
condition 
\begin{eqnarray*}
\delta u_\tau (x) = 0, 
\end{eqnarray*} 
is the non-linear diffusion equation
\footnote{While in the case of a manifold without boundary, the viscosity term $\nu \triangle_1u_\tau$ ($\nu$ is the kinematical viscosity)
commutes
with $P$, in the case with boundary we have to impose it from the beginning, i.e. we take $\nu P\triangle_1u_\tau$ (c.f. page 144 in [17]) and in
any case we are left with the expression in the r.h.s. of eq. $(83)$ or eq. $(84)$.}
\begin{eqnarray}
{\partial u \over \partial \tau}(\tau,x) + {\cal P} \nabla^g_{u_\tau^g} u_\tau= \nu {\cal P}\triangle_1 u_\tau 
\end{eqnarray}
where ${\cal P}$ denotes the projection operator into the co-closed component of the Helmholtz-Hodge decomposition
of $u_\tau$ ($\tau \geq 0$)
which in view of the trivial identity (c.f. theorem $1.17$ in [ 16 ])
\begin{eqnarray*}
L_{u_\tau^g}u_\tau = \nabla^g_{u_\tau^g}u_\tau + {1\over 2}d(g(u_\tau,u_\tau))),
\end{eqnarray*}
so that the non-linear term is
\begin{eqnarray*}
{\cal P}\nabla^g_{u^g_\tau}u_\tau = {\cal P}L_{u^g_\tau}u_\tau,
\end{eqnarray*}
since by definition ${\cal P}$ vanishes on exact $1$-forms, and thus the Navier-Stokes equation takes the form
\begin{eqnarray*}
{\partial u \over \partial \tau} = {\cal P}(\triangle _1 u_\tau + L_{-u_\tau^g}u_\tau)
\end{eqnarray*}
which from expression $(10)$ we readily conclude that it can be rewritten as [4-6,12,13]
\begin{eqnarray}
{\partial u \over \partial \tau}(\tau,x) = {\cal P}H_1(2\nu g,-{1 \over 2\nu}u_\tau)u_\tau (x), x\in M, \tau \geq 0.
\end{eqnarray}
We further have the no-slip boundary condition given by
\begin{eqnarray*}
u|_{\partial M} = 0. 
\end{eqnarray*}

If we know define the vorticity $2$-form of the fluid as 
\begin{eqnarray*}
\Omega= \Omega_\tau(x) = du_\tau (x),
\end{eqnarray*}
 for any $x \in M, \tau \geq 0$.  Let us apply
$d$ to $(84)$; in account that in writing down the Hodge decomposition of the non-linear term 
we obtain that
\begin{eqnarray*}
d{\cal P}L_{- u^g_\tau}u_\tau= L_{- u^g_\tau}du_\tau = L_{- u^g_\tau}\Omega_\tau,
\end{eqnarray*}
and that similarly, further taking in account $(11)$ we obtain that 
\begin{eqnarray*}
d{\cal P}\triangle_1 u_\tau = \triangle_2 \Omega_\tau,
\end{eqnarray*}
so that altogether we obtain
the Navier-Stokes equations for the vorticity 
\begin{eqnarray}
{\partial \Omega \over \partial \tau}(\tau,x) = H_2(2\nu g,-{1 \over 2\nu}u_\tau)\Omega_\tau (x), x\in M, 0\leq \tau.
\end{eqnarray}
Since we have assumed that $M$ has a smooth boundary $\partial M$, our interest
now resides in boundary conditions.
Then, 
\begin{eqnarray*}
\Omega = {1 \over 2} 
\Omega_{\alpha \beta}dx^\alpha \wedge dx^\beta
\end{eqnarray*}
where for simplicity we have ommited the variables $(\tau,x)$,
satisfies the absolute boundary conditions iff
\begin{eqnarray}
\Omega_{\alpha n} = 0, \forall {\alpha= 1 \ldots n-1}, ~ {\rm and}~ {\partial \over \partial x^n}\Omega_{\alpha \beta} = 0,1\leq \alpha <\beta\leq n.
\end{eqnarray}
These are the boundary conditions considered in Fluid Mechanics for the vorticity 
\footnote{Indeed, from the no-slip boundary condition, it follows say, in the flat
$3d$ case that the vector {\it normal}
 to $\partial M$ given by $\Omega_\tau = {\rm rot~}u_\tau$ (for every $\tau$) vanishes, which in invariant form is
nothing else than $\Omega_{\alpha n} = 0$, for every $\alpha = 1,\ldots, n-1$. 
The condition $(d\Omega_\tau)_{\rm norm} = 0, i.e. {\partial \Omega \over
\partial n} =0$, represents that the flow of vorticity through the normal to the boundary is zero, which is a natural condition for a non-permeable boundary,
 for every $\tau$.}.
Now if we apply $\delta $ to the definition of the vorticity, from eqt. $(11)$ we obtain that 
\begin{eqnarray*}
\delta \Omega_\tau = \delta du_\tau = -\triangle_1 u_\tau + d\delta u_\tau
= - 2 H_1(g,0)u_\tau, 
\end{eqnarray*}
where in the last identity we have taken in account the incompressibility condition and eqt. $(10)$ with $Q \equiv 0$, and thus 
we obtain the Poisson-de Rham equation for the velocity
\begin{eqnarray}
H_1(g,0)u_\tau (x)= -{1\over 2}\delta \Omega_\tau (x), \tau \geq 0, x\in M.
\end{eqnarray}
We have already described in [5,6 \& 13], the representations for this equation in the case of smooth boundary, by running either a gradient
diffusion  process by isometrically embedding $M$ in an Euclidean space, or by running random Hessian and Ricci flows in the case of an arbitrary
compact manifold, to solve for the vorticity and the velocity, respectively. Yet, in the above mentioned articles, the representation for the vorticity in the smooth boundary case was not given, so we shall
now deal with this problem. 

We start by noting that from eqts. $(1, 10 \& 85)$ and our constructions in Section $I$ we can conclude that there is a Riemann-Cartan-Weyl connection on $TM$ associated to the Navier-Stokes operator. This connection is determined by the
Riemannian metric $2\nu g$, it is compatible with this metric, and has a trace-torsion determined by the time-dependant
$1$-form $Q(\tau,x)= Q_\tau(x)= {-1 \over 2\nu}u(\tau,x)$ ($x \in M, \tau \ge 0$) [4-6,13]. Thus, this "Navier-Stokes" connection, which 
we shall denote as $\nabla$, has for Christoffel coefficients, the $n^3$ time-dependant functions on $M$
defined by \footnote{The first term in $(88)$ designate the Christoffel coefficients of the Levi-Civita connection $\nabla^g$ defined in terms of
the metric and its first order derivatives.}
\begin{eqnarray}
\Gamma ^{\alpha}_{\beta \gamma}(\tau,x) = 2\nu{\alpha\brace\beta\gamma}(x)~+~ {2 \over (n-1)}
\left\{{-1 \over 2\nu}\delta^\alpha_\beta ~u(\tau,x)_\gamma ~ + ~{ 1 \over 2 \nu} g_{\beta\gamma}(x)~u(\tau,x)^\alpha\right\},
\end{eqnarray} 
so that we have a time-dependant horizontal canonical vector field, $L^\nabla (\tau,r) =(L^\nabla_a(\tau,r)),a =1,\ldots , n)$ associated to $\nabla$ by $(4)$.
It is trivial to check that expression $(88)$ indeed defines a connection compatible with the metric $2\nu g$ and with trace-torsion given by ${- 1 \over 2\nu}u_\tau$.
As long as eqt. $(84)$ admits an unique solution in $[0,T]$, then this connection is uniquely determined in $[0,T]$ as well.

We wish to solve the boundary-initial-value problem for 
\begin{eqnarray*}
\tilde \Omega \in \Lambda^2([0,T]\times P_{O(n)}), {\rm for}~ T>0
\end{eqnarray*}
given by the Navier-Stokes equation for the vorticity as written on $P_{O(n)}$:
\begin{eqnarray}
{\partial \tilde \Omega \over \partial \tau}(\tau,r) = \tilde H_2(2\nu g,-{1 \over 2\nu}u_\tau)\tilde\Omega_\tau (r),
~~~\tilde\Omega(0,r) = \tilde \omega(r)
\end{eqnarray}
where $\tilde \omega$ is the horizontal lift of a given $2$-form $\omega$ defined on $M$ (the vorticity at time $0$), and \footnote{These are the expressions
on $P_{O(n)}$ of the given absolute boundary conditions which in p.d.e. theory are "Robin"-type conditions.}
\begin{eqnarray}
\theta^b_\beta(\theta^a_n \Omega_{ab} + \theta^a_\alpha {\partial \tilde \Omega_{ab}\over
\partial x^n})(\tau,r) = 0, ~{\rm on~}\partial P_{O(n)},\forall \tau\geq 0,\alpha,b=1,\ldots,n-1.
\end{eqnarray} 
From $(35)$ we note that we can rewrite the initial-value problem $(89)$ as
\begin{eqnarray}
{\partial \tilde \Omega \over \partial \tau}(\tau,r)& 
=& \nu (L_a^{\nabla} (\tau,r))(L_a^{\nabla}) (\tau,r))(\tilde \Omega_\tau(r)) 
+ \nu {\rm Ric}^\flat \otimes\tilde \Omega(\tau,r)\nonumber\\& - & 2\nu \tilde R\otimes\tilde \Omega (\tau,r), 0\le \tau \le T,
\end{eqnarray}
with
\begin{eqnarray}
\tilde \Omega (0) = \tilde \omega,
\end{eqnarray}
and we already proved (c.f. equation $(45)$) that the boundary conditions admit the expression

\begin{eqnarray*}
P\otimes \theta \otimes \tilde \Omega^\flat + Q\otimes \theta \otimes [{\partial \tilde \Omega^\flat _\tau
\over \partial x^n} -
 {\partial \tilde \Omega^\flat_\tau \over \partial e^\alpha_m} \otimes \Gamma^\alpha _{n\beta}(\tau,x)\otimes e^\beta_m(x)
\end{eqnarray*}
\begin{eqnarray}
 +
 e(x)\otimes \Gamma_n(\tau,x)\otimes \theta (x)\otimes \tilde \Omega^\flat ](\tau,r) = 0, \forall r =(x,e) \in \partial P_{O(n)},
\end{eqnarray}
where $\Gamma_n(\tau,x) = (\Gamma^\alpha_{n\beta}(\tau,x))$, and $\tilde \Omega^\flat _\tau$ (for $0 \le \tau \leq T$)
is in $L(TM, R^{n*})$ and verifies that its partial horizontal lift $e(x)\otimes \tilde \Omega_\tau (r)$
is the corresponding time-dependant $2$-form on $P_{O(n)}$ in the boundary-initial-value 
problem $(91-93)$. The essential difference with $(41 \& 45)$ is that the differential operator has a time-dependant drift term.

Consider the Stratonovich stochastic differential equation for the diffusion
process $r(r,\tau,t)= (X(x,\tau,t),e(e,\tau,t))$ on $P_{O(n)}\sim R^n_+\times R^{n^2}$ defined by running backwards in time the s.d.e. $(47)$; i.e. :
\begin{eqnarray}
dr^\alpha (r,\tau,t) & = & L^\nabla_k(\tau  - t, r(r,\tau,t))\circ dB^k(t)\nonumber\\
 &+ & \delta ^{\alpha}_nd\phi (\tau -t),  0\leq t \leq \tau \leq T
\end{eqnarray}
$(\alpha =1,\ldots,n$) with initial condition
\begin{eqnarray}
r(r,\tau,0) = r, 0\leq t\leq \tau \leq T.
\end{eqnarray}
 We note that it can be
rewritten as
\begin{eqnarray}
dX^\alpha (x,\tau,t) = e^\alpha_k (e,\tau,t) \circ dB^k(t) + \delta ^{\alpha}_nd\phi (\tau -t),  0\leq t \leq \tau \leq T
\end{eqnarray}
\begin{eqnarray*}
de^\alpha_ k(e,\tau,t) = 
-\Gamma^\alpha _{\beta \gamma}(\tau-t,X(x,\tau,t))e^\gamma_k(e,\tau,t) \circ dX^\beta(x,\tau,t) 
\end{eqnarray*}
\begin{eqnarray*}
 = - \Gamma^\alpha _{\beta \gamma}(\tau -t,X(x,\tau,t))e^\gamma_k(e,\tau,t)e^\beta_p(e,\tau, t)\circ dB^p(t) 
\end{eqnarray*}
\begin{eqnarray}
-\Gamma^\alpha _{n \gamma}(\tau -t,X(x,\tau,t))e^\gamma_k(e,\tau,t)d\phi (\tau -t), 0\leq t \leq \tau \leq T,
\end{eqnarray}
($\alpha,k =1,\ldots,n$) with initial conditions
\begin{eqnarray}
X(x,\tau,0) = x, e^\alpha (e,\tau,0)= e(x) \in T_xM, 0 \leq t \leq \tau \leq T.
\end{eqnarray}
Here, $\phi(\tau)$ is a continuous non-decreasing process which increases only when $X(x,\tau, t) \in \partial M$. \footnote{Here
we are assuming that $B$ and $\phi$ satisfy eqts. $55 \& 56$.}. Having assumed that $g_{\alpha \beta} \in C^\infty_b (M)$, if we further assume
that $u(\tau,x)$ has all its components $u(\tau,x)_\alpha \in C^\infty_b (M)$ ($1\le \alpha \le n$) 
and that further $\phi \in C^2_b (M)$, then $(96-98)$ has a strong unique solution.

For $\tilde \omega\in C_0(P_{O(n)}\rightarrow L(TM,R^{n^*}))$ and $\tau \in [0,T]$, set
\begin{eqnarray}
(H_\tau \tilde \omega)(r) : =  E_r[M(\tau,\tau,w)\otimes \tilde \omega(r(r,\tau,\tau,w))]  .
\end{eqnarray}
where $M(\tau,t) =(M^a_\alpha (\tau,t,w))$ is a $R^ n \otimes R^{n^*}$-valued process given by
\begin{eqnarray}
M (\tau,t,w)= K(\tau,t,w)\otimes \theta (e,\tau,t), 0 \leq t \leq \tau \leq T,
\end{eqnarray}
i.e. componentwise
\begin{eqnarray*}
M^a_\alpha (\tau,t,w) =
K_\alpha^\beta (\tau,t,w)\theta_\beta^a (\tau,t,w),
\end{eqnarray*}
where $K(\tau,t,w) = (K^{\alpha \beta}_\gamma(\tau,t,w))$ is the unique solution of the s.d.e defined as follows: Set
$K^1(\tau,t,w) = K(\tau,t,w)\otimes P$ and $K^2(\tau,t,w) = K(\tau,t,w)\otimes Q$ so that $K(\tau,t,w) = K^1(\tau,t,w) + K^2(\tau,t,w)$,
where for any $\tau \in [0,T]$ such that  $X(x,\tau,t) \in {\rm int}(M)$, $t \in [0,\tau]$
 we have \footnote{We shall omit in the following the variable $w$, as customary, in the expression for
$K$ and its specialization to the flat case.}
\begin{eqnarray}
(i)~~dK^1(\tau,t) :& = & dK(\tau,t)\otimes P = K(\tau,t)\otimes [\theta(e,\tau,t)\otimes de(e,\tau,t)\nonumber\\
& +& \{\nu{\rm Ric}(X(x,\tau,t)) -2\nu R(X(x,\tau,t))\}dt ]\otimes P,
\end{eqnarray}
and for any $\tau \in [0,T]$,
\begin{eqnarray*}
(ii)~~dK^2(\tau,t): & = & dK^2(\tau,t)\otimes Q\nonumber\\
& = & K(\tau,t)\otimes [\theta(e,\tau,t)\otimes de(e,\tau,t)
\end{eqnarray*}
\begin{eqnarray} 
+ \{\nu{\rm Ric}(X(x,\tau,t)) - 2 \nu  R(X(x,\tau,t))\}dt ]  I_{{\rm int(M)}}(X(x,\tau,t))\otimes Q,
\end{eqnarray}
and with probability $1$, $\tau \rightarrow K^1(\tau ,t)$ is right-continuous with left-hand limits: furthermore
\begin{eqnarray}
K^1(\tau,t) = 0, {\rm if}~ X(x,\tau,t) \in \{x^n=0\},
\end{eqnarray}
and the initial value is
\begin{eqnarray}
K^1(\tau,0) = I_{{\rm int~(M)}} (X(x,\tau,0))e(e,\tau,0)\otimes P, K^2(\tau,0) = e(e,\tau,0)\otimes Q.
\end{eqnarray}

{\bf Theorem  7.} Let $\tilde \omega = \tilde \omega (r)$ a be function in $C_0(P_{O(n)} \rightarrow L(TM, R^{n^*}))$. Then, 
$(H_\tau \tilde \omega)(r) = E_r[M(\tau,\tau,w)\otimes \tilde \omega(r(\tau,\tau,w))]$ is the unique solution of $(91)$, with initial value 
$\tilde \Omega (0,-) = \tilde \omega(-)$, and further satisfying the boundary condition given by $(93)$.

{\bf Proof:  } It follows easily from observing that this is the time-dependant version of Theorem $5$.

Now, let $\Phi (\tau,x) = \Phi_\tau(x) \in \Lambda^ 2([0,T] \times T^*M)$ such that its horizontal (partial) lift 
$\Phi_\tau (x) \otimes e(x)=(H_\tau \tilde \omega)(r)$, for all $r=(x,e(x))$, $0 \leq \tau \leq T$. 
Then, from Theorem $7$ follows that $\Phi$ is the unique solution of the boundary-initial-value problem
\begin{eqnarray}
{\partial \Omega \over \partial \tau}(\tau,x) = H_2(2\nu g,{- 1 \over 2\nu}u_\tau)\Omega_\tau (x), ~~~~\Omega_0(x) = \omega
(x),0\leq \tau < 0, x \in M,
\end{eqnarray}
where $\omega$ is the $2$-form on $M$ such that its horizontal lift is $\tilde \omega$, i.e. $\omega (x)\otimes e (x)\otimes e(x) =\tilde \omega (r)$, 
and satisfying further the absolute boundary conditions:
\begin{eqnarray}
\Omega_{\alpha n} = 0,{\partial \over \partial x^n}\Omega_{\alpha \beta} = 0, 1\leq \alpha <\beta \leq n-1.
\end{eqnarray}
Thus, we have obtained the representation for the Navier-Stokes equations for the vorticity with absolute boundary conditions. 

Now we can solve for the vorticity equation in the flat space $R^n_+$. Now since in this case
$g = I$, the identity, and thus in $(88)$ the expression for the Christoffel symbols reduce to the form
\begin{eqnarray}
\Gamma ^{\alpha}_{\beta \gamma}(\tau,x) = {2 \over (n-1)}
\left\{{-1 \over 2\nu}\delta^\alpha_\beta ~u(\tau,x)_\gamma ~ + {1\over  2\nu} \delta_{\beta\gamma}~u(\tau,x)^\alpha\right\},
\end{eqnarray} 
so we now consider the s.d.e.
\begin{eqnarray}
dX^\alpha (x,\tau,t) = e^\alpha_k (e,\tau, t) \circ dB^k(t) + \delta ^{\alpha}_nd\phi (\tau -t), 0\leq t \leq \tau,
\end{eqnarray}
\begin{eqnarray}
de^\alpha_ k(e,\tau,t) &= &
 = - \Gamma^\alpha _{\beta \gamma}(\tau -t,X(x,\tau,t))e^\gamma_k(e,\tau , t)e^\beta_p(r,\tau, t)\circ dB^p(t)\nonumber\\
& - &\Gamma^\alpha _{n \gamma}(\tau -t,X(x,\tau,t))e^\gamma_k(e,\tau, t)d\phi (\tau -t), 
\end{eqnarray}
($\alpha,k=1,\ldots,n$) with initial conditions
\begin{eqnarray}
X(x,\tau,0) = x, e^\alpha (e,\tau,0)= e(x) \in T_xM, 0 \leq t \leq \tau.
\end{eqnarray}
Here, $\phi(\tau)$ is a continuous non-decreasing process which increases only when $X(\tau, t,x) \in \{x^n = 0\}$.
 Furthermore, since ${\rm Ric}$ and $R$ are identically equal to $0$ in
the expressions $(101 \& 102)$, so that $K(\tau,t)= K^1(\tau,t) + K^2(\tau,t)$ is the solution of the problem described as follows: for any
$\tau \in [0,T], x \in \partial M$ we have
\begin{eqnarray}
dK^1(\tau,t) : =  dK(\tau,t)\otimes P = K(\tau,t)\otimes \theta(\tau,t)\otimes de(\tau,t)\otimes P,
\end{eqnarray}
and for any $\tau \in [0,T]$,
\begin{eqnarray}
dK^2(\tau,t): =
K(\tau,t)\otimes \theta(\tau,t)\otimes de(\tau,t)I_{{\rm int(M)}}(X(\tau,t))\otimes Q,
\end{eqnarray}
satisfying
\begin{eqnarray}
K^1(\tau,t) = 0, {\rm if}~ X(\tau,t) \in \{x^n = 0\},
\end{eqnarray}
and the initial value is
\begin{eqnarray}
K^1(\tau,0) = I_{{\rm int(M)}}(X(\tau,0))e(\tau,0)\otimes P, K^2(\tau,0) = e(\tau,0)\otimes Q.
\end{eqnarray}
Thus, if we finally consider $M(\tau,t,w) = K(\tau,t,w)\otimes \theta(\tau,t,w)$ (where $\theta = e^{-1}$)
for $\tilde \omega\in C_0(P_{O(n)}\rightarrow L(R^n,R^{n^*}))$ and $0\leq \tau \le T$, set
\begin{eqnarray}
(H_\tau \tilde \omega)(r) := E_r[M(\tau,\tau,w)\otimes \tilde \omega(r(r,\tau,\tau,w))].
\end{eqnarray}

{\bf Theorem 8.} The $2$-form $\Phi _\tau \in \Lambda^2 ([0,T]\times R^{n^*})$ such that
\begin{eqnarray}
\Phi_\tau (x)\otimes e(x)=(H_\tau \tilde \omega)(r), r=(x,e(x)), 
\end{eqnarray}
is the unique solution of the boundary-initial-value problem given by 
\begin{eqnarray}
{\partial  \Omega \over \partial \tau} = H_2(2\nu I,-{1 \over 2\nu}u_\tau)
\Omega_\tau \equiv \nu {\rm div~grad~}\Omega_\tau 
- L_{u_\tau}\Omega_\tau
\end{eqnarray}
with initial value
\begin{eqnarray}
e(0)\otimes \Phi(0,-) = \tilde \omega(-)
\end{eqnarray}
and further satisfying the absolute boundary conditions .

\section{Kinematic Dynamo of Magnetohydrodynamics on Smooth Manifolds
 with Smooth boundary}
\hspace{2em}
We can extend these constructions to solve the passive transport equations of magnetohydrodynamics. All
 we have to do is to construct the multiplicative horizontal operator functional for the differential operator
$H_{n-1}(2\nu^m g,{-1 \over 2\nu^m}u_\tau)$, with $\nu^m$ the magnetic diffusivity, $u_\tau$ given by the solution of either the
Euler equations, or still the Navier-Stokes equations, with absolute boundary conditions. Indeed, doing this
(for which we have to take in account the coupling of the 
Riemannian curvature to the $(n-1)$-magnetic form in the Weitzenbock term for the definition of $K$), 
we can then solve the kinematical dynamo problem for a magnetic field $B(\tau,x)$ defined by
duality as 
\begin{eqnarray}
i_{B_\tau}\mu =  \Omega_\tau, 0 \leq \tau \le T
\end{eqnarray}
where $i$ denotes the interior product derivation, 
$\mu = {\rm det (g)}^{1\over 2}dx^1\wedge \ldots \wedge dx^n$ 
is the Riemannian volume density, and
$\Omega (\tau,x) \in \Lambda^{n-1}([0,T]\times T^*M)$ satisfies the transport equations 
(kinematic dynamo problem)
\begin{eqnarray}
{\partial \Omega \over \partial \tau} = 
H_{n-1}(2\nu^m g,{-1 \over 2\nu^m}u_\tau) \Omega_\tau, \tilde \Omega_0 =i_{B_0}\mu,
0 \leq \tau <T, 
\end{eqnarray}
for given
\begin{eqnarray}
B(0,-) = B(-),
\end{eqnarray}
satisfying the absolute boundary conditions for $ \Omega$.

\section{Conclusions}
In this article, we have given random representations for the Navier-
Stokes equations for the vorticity, and the kinematic dynamo equation of
magnetohydrodymamics, for
incompressible fluids on smooth boundary manifolds, and in particular, in the case of
flat euclidean space. No such general representations were known but for the case of empty 
boundary smooth compact manifolds [5,12,13], and as an implementation
of the case of manifolds isometrically immersed in Euclidean space, for Euclidean
space itself. We would like to remark that this program stemmed as a covariant
extension of the random vortex method in 2D (without boundary)
of Computational Fluid Dynamics ( see A. Chorin [32], and references therein).
In the latter method,due to the fact that in flat 2D the vorticity can be identified with a scalar
field, the vorticity equation can be integrated by using the Ito formula for scalar fields.
The extension of this formula, has been the backbone for the obtention of our representations in both the
empty and non-empty boundary cases. We should also stress that product formulas (such as those arising from transition densities as is the case for diffusion equations)
have been algorithmically 
implemented  in Fluid Dynamics for a long period [33]. Yet, more specifically related to the present approach,
numerical methods for the random
 integration of nonlinear partial differential equations have been developed (c.f.
[34]); furthermore these methods have been implemented for p.d.e.'s for scalar fields satisfying reflecting boundary conditions [35-38].  
Thus we might  expect that the former approach properly extended to differential forms and implemented for the
representations achieved in this article, will eventually lead to interesting numerics.


\begin{thebibliography}{99}
 
\bibitem{ } N.Ikeda \&S. Watanabe, {\bf Stochastic Differential
Equations on Manifolds}, (North-Holland/Kodansha, Amsterdam/Tokyo, 1981). 

\bibitem{  } K.D. Elworthy,{\bf Stochastic Differential Equations on Manifolds}, (Cambridge Univ.
Press, Cambridge, 1982).

\bibitem{  } H. Airault, {\it Perturbations singulieres et solutions stochastiques de problemes de
D.Neumann-Spencer, J.of Pure and Appl.Math.} {\bf 55} (1976),233-268. 

\bibitem{  } D. Rapoport, {\it Torsion and Quantum, thermodynamical and hydrodynamical
fluctuations}, in  {\bf The Eighth Marcel Grossmann Meeting in General Relativity, Gravitation
and Field Theory, Proceedings, Jerusalem, June 1997},p. 73-76, vol. edts. A, T.Piran and R.Ruffini
, (World Scientific, Singapore, 1999).

\bibitem{  } D. Rapoport, {\it Random representations for viscous fluids and the passive magnetic fields transported on them}, in {\bf
 Proceedings of the Third International Conference on Differential Equations and Dynamical Systems, Atlanta, May 2000, special issue,  Discrete 
and Continuous Dynamical Systems, series B}, 
ed. S. Hu , 2000, (2001), 327-336.

\bibitem{  } D. Rapoport,  math-ph/0012032, preprint IN000-33GTF, Geometry and Topology
of Fluid Flows Program, Isaac Newton Institute for Mathematical Sciences, 
Univ.of Cambridge, December 2000.

\bibitem{  } P. Malliavin, {\it Formule de la moyenne, calcul de pertubations et th\'eoremes
d'annulation pour les formes harmoniques,  Journal of Functional Analysis}, {\bf 17} (1974), 276-291.
   
\bibitem{  } P. Malliavin,  {\bf G\'eom\'etrie Differentielle
Stochastique}, (Les Presses Univ. Montreal ,1978).

\bibitem{  } L. Rogers \& D. Williams, {\bf Diffusions, Markov Processes and Martingales, vol. II},
(John Wiley, New York, 1989).

\bibitem{  } M.Vishik \& A.Fursikov,  {\bf Mathematical Problems of Statistical
Hydrodynamics}, (Kluwer Academic Press, Dordrecht, 1989).
 
\bibitem{  } A. Monin  \& A.Yaglom, {\bf Statistical Fluid Mechanics, vol. 
II}, ed. J. Lumley, (M.I.T. Press, Cambridge (MA), 1975).

\bibitem{  } D. Rapoport,  {\it Random Geometry of Quantum Mechanics, Relativity and Fluid-Dynamics},
in {\bf Open Problems of Science at the End of the Millenium, vol. II, p.243-276, Proceedings of the
Conference on Fundamental Problems in Science at the End of the Millenium, Beijing, September 1998}, T. Gill et al. (edts.), (Hadronic Press, Palm Harbor (USA), 1999).

\bibitem{   }D. Rapoport,  {\it Random diffeomorphims and integration of the classical Navier-Stokes equations, Reports in Mathematical Physics}{\bf 49}
, no.1,(2002) pp.1-27.

\bibitem{   } Ya.Belopolskaya \& Yu. Dalecki,  {\bf Stochastic Processes and Differential Geometry},
(Kluwer Academic Press, Dordrecht, 1989).

\bibitem{   } V. Arnold,  {\it Sur la g\'eometrie diff\'erentielle des groupes de Lie de dimension infinie
et ses applications a l'hydrodynamique des fluides parfaits,  Ann. Inst. Fourier} {\bf 16}, (1966), 319-361.

\bibitem{  } V.Arnold \& B.Khezin, {\bf Topological Methods in Hydrodynamics}, (Springer Verlag Series
in Applied Mathematical 
Sciences 125, New York/Berlin, 1999).

\bibitem{  } D. Ebin \&  J. Marsden, {\it Groups of diffeomorphisms and the motion of an incompressible fluid, Annals of
Mathematics} {\bf 92}, (1970),102-163.

\bibitem{   } Yu. Gliklikh, {\bf Global Analysis in Mathematical Physics (Geometric and Stochastic Methods)}
, (Springer Verlag Applied Mathematical Sciences 122, New-York/Berlin, 1997).

\bibitem{   } Yu. Gliklikh, {\bf Ordinary and Stochastic Differential Geometry as a Tool for Mathematical-
Physics}, (Kluwer, Dordrecht, 1996).

\bibitem{   } Yu. Gliklikh,  {\it Viscous Hydrodynamics through stochastic perturbations of flow of
perfect fluids on groups of diffeomorphims, Proceedings of the Voronezh State University},{\bf 1},
2001, p.83-91.

\bibitem{   } D.Rapoport, {\it Covariant Thermodynamics and the Ergodic Theory of Quantum and Thermodynamical 
Flows}, in {\bf Instabilities and Nonequilibrium Structures, vol. VI,  Proceedings, VIth. International
Workshop on Instabilities and Nonequilibrium Structures, Valparaiso (Chile), Dec. 1995}, 359-370,
edts. E. Tirapegui et al,
(Kluwer Series in Complex Systems, Dordrecht, 2000).
\bibitem{   } D.Rapoport, \& S.Sternberg, {\it On the interactions of spin with torsion,  Annals of Physics}
{\bf vol. 158}, (1984), 447-475.
\bibitem{  } D. Rapoport, {\it Torsion, Brownian Motion, Quantum Mechanics and Fluid-dynamics I \& II}, in
{\bf Proceedings of the Ninth International Marcel Grossman Meeting in Relativity, Gravitation and Field Theory, vol. III (Univ. of Rome, June 2000,
edts. R.Ruffini et al}, (World Scientific, Singapore, 2002), and www.icra.it/MG/mg9/Proceedings/Proceedings.html  ;


\bibitem{  } D. Rapoport, {\it Torsion and Nonlinear Quantum Mechanics}, in {\bf Group XXI, Physical Applications of
Aspects of Geometry, Groups and Algebras, Proceedings, XXI International Conference on Group
Theoretical Methods in Physics, Goslar (Germany), June 1995}, edts. H. Doebner et al, (World
Scientific, Singapore, 1997); ibid.
Riemann-Cartan-Weyl Quantum Geometries and the equivalence of
the Maxwell and Dirac-Hestenes equations, {\it Advances in Clifford Algebras and its Applications} {\bf vol. 8}. No.1,
p. 126-149 (1998).

\bibitem{  } D. Rapoport, {\it The Geometry of Quantum Fluctuations, I \& II}, in {\bf New Frontiers of
Algebras, Groups and Geometries, Proceedings, International Conference on the New Frontiers
of Algebras, Groups and Geometries, Monteroduni (Italy), August 1995}, ed. G. Tsagas, (Hadronic Press,
Palm Harbor, Florida, USA, 1996).

\bibitem{   } D. Rapoport,  {\it The Geometry of Quantum Fluctuations, the Quantum Lyapunov exponents
and the Perron-Frobenius Semigroups}, in {\bf Dynamical Systems and Chaos, vol. II,
Proceedings,
International Conference on Dynamical Systems and Chaos, Tokyo (May 1994)}, pags. 73-77, ed. Y. Aizawa,
(World Scientific, Singapore ,1995).

\bibitem{  } D. Rapoport, {\it The Cartan-Weyl Stochastic Processes of Gravitation,  Int. Journal of
Theoretical Physics} {\bf vol. 30}, No.11, (1991), 287-310; ibid {\it Riemann-Cartan-Weyl Quantum Geometry II, Int. J.Theor.Phys.}{\bf 36}, (1997)
 no.10, 2115-2152; ibid.
{it Riemann-Cartan-Weyl Quantum Geometry I, Int. J. Theor.Phys.}{\bf 35}, no.2, (1996), 287-309. 
 
\bibitem{  } D. Rapoport,\& S. Sternberg, {\it Classical Mechanics without lagrangians nor hamiltoneans, Nuovo Cimento} {\bf 80}, (1984), 371-383.
\bibitem{  } D. Rapoport,  {\it Cartan Structure of Classical and Quantum Gravity}, in {\bf Gravitation, The Space-time
Structure, Proceedings, Latinoamerican Symposium of Relativity and Gravitation, Aguas de Lindoia (Brasil), June
1993}, edts. W. Rodrigues and P.Letelier , ( World Scientific, Singapore, 1995), 220-229.


\bibitem{  } P.Malliavin, {\bf Stochastic Analysis}, (Springer Verlag, Berlin, 1997).

\bibitem{   } B.Dubrovin, A. Fomenko \& S.Novikov,  {\bf Modern Geometry-Methods and Applications, vol.I},
(Springer Verlag, Berlin, 1995).

\bibitem{   } A. Chorin, {\bf Turbulence and Vorticity}, (Springer Verlag Series in Applied Mathematics, Springer
Verlag, Berlin, 1995).

\bibitem{  } A. Chorin, T.J.R. Hughes, J.Marsden \& M. McCracken, {\it Product formulas and Numerical
Algorithms,  Comm.Pure and Applied Math.}, {\bf 31}, (1978), 205-256.

\bibitem{  } D.Talay and L.Tubaro, {\it Probabilistic Models for Nonlinear Partial Differential Equations, C.I.M.E. Lectures
(May 22-30)}, {\bf Lecture Notes in Mathematics vol. 1507},( Springer Verlag ,1996).

\bibitem{  } L.Slominsky, {\it On existance, uniqueness and stability of solutions of multidimensional s.d.e.'s with reflecting boundary
conditions,  Ann.I.H.Poincar\'e} {\bf 29} (1993).

\bibitem{  } L.Slominsky, {\it On approximation of solutions of multidimensional s.d.e.'s with reflecting boundary conditions,
Stochastic processes and their Applications}, {\bf 50}(2), (1994),197-219.

\bibitem{  }D. L\'eplinge, {\it Euler scheme for reflected stochastic differential equations,  Mathematics and Computers in
Simulation}, vol. {\bf 38} (1995).
\bibitem{  } D.L\'eplinge, {\it Un sch\'ema d'Euler pur \'equations diff\'erentielles r\'efl'echies, Note aux Comptes Rendus de l'
Acad\'emie des Sciences}, {\bf 316}(1993) 601-605.



\bibitem{  } A. Chorin, {\bf Turbulence and  Vorticity}, (Springer Verlag Series in Applied Mathematics, Berlin, 1995).

\bibitem{  } D. Stroock \& S.R.S. Varadhan,  {\bf Multidimensional Diffusion Processes},
(Springer, Berlin,1989). 


\end{thebibliography}
\end{document}